\documentclass[aps,nofootinbib,prd,superscriptaddress,tightenlines,notitlepage,
twocolumn,showpacs,floatfix]{revtex4-1}
\usepackage[colorlinks]{hyperref}
\usepackage{tabularx}
\usepackage{graphicx}
\usepackage{amssymb,bm,tensor}
\usepackage{xcolor}
\usepackage{amsmath}
\usepackage[varg]{txfonts}
\usepackage{enumerate}
\usepackage[normalem]{ulem}
\usepackage{bm}
\usepackage[OT1]{fontenc}
\usepackage{orcidlink}

\begin{document}

\title{Effects of orbital eccentricity on  continuous gravitational waveforms from triaxially deformed precessing neutron stars in tight binaries}

\author{Wen-Fan Feng \orcidlink{0000-0001-5033-6168}}
\email{fengwf@pku.edu.cn}
\affiliation{Kavli Institute for Astronomy and Astrophysics, Peking University, Beijing 100871, China}

\author{Tan Liu \orcidlink{0000-0002-2898-1360}}
\email{lewton@mail.ustc.edu.cn}
\affiliation{School of Fundamental Physics and Mathematical Sciences, Hangzhou Institute for Advanced Study, UCAS, Hangzhou 310024, China}
\affiliation{University of Chinese Academy of Sciences, 100049/100190 Beijing, China}

\author{Yan Wang \orcidlink{0000-0001-8990-5700}}
\email{ywang12@hust.edu.cn}
\affiliation{MOE Key Laboratory of Fundamental Physical Quantities Measurements, PGMF, Department of Astronomy and School of Physics, Huazhong University of Science and Technology, Wuhan 430074, China} 

\author{Lijing Shao \orcidlink{0000-0002-1334-8853}}
\email{lshao@pku.edu.cn}
\affiliation{Kavli Institute for Astronomy and Astrophysics, Peking University, Beijing 100871, China}%
\affiliation{National Astronomical Observatories, Chinese Academy of Sciences, Beijing 100012, China}

\date{\today}

\begin{abstract}

The successful detection of continuous gravitational waves from spinning neutron stars (NSs) will shape our understanding of the physical properties of dense matter under extreme conditions. 
Binary population synthesis simulations show that forthcoming space-borne gravitational wave detectors may be capable of detecting some tight Galactic double NSs with 10-min orbital periods. 
Successfully searching for continuous waves from the individual NS in such a close binary demands extremely precise waveform templates considering the interaction between the NS and its companion.
Unlike the isolated formation channel, double NS systems from the dynamical formation channel have moderate to high orbital eccentricities.
To accommodate these systems, we generalize the analytical waveforms from triaxial nonaligned NSs under spin-orbit coupling 
derived by Feng \textit{et al.} [\href{https://journals.aps.org/prd/abstract/10.1103/PhysRevD.108.063035}{Phys. Rev. D 108, 063035 (2023)}] 
to incorporate the effects of the orbital eccentricity. 
Our findings suggest that for binaries formed through isolated binary evolution, the impact of eccentricity on the continuous waves of their NSs can be neglected.
In contrast, for those formed through dynamical processes, it is necessary to consider eccentricity, as high-eccentricity orbits can result in a fitting factor of $\lesssim 0.97$ (0.9) within approximately 0.5 (1) to 2 (5) yr of a coherent search (at wave frequencies of 100 and 200~Hz).
Once the continuous waves from spinning NSs in tight binaries are detected, the relative measurement accuracy of eccentricity can reach $\Delta e / e \sim O(10^{-7})$ for a signal-to-noise ratio of $O(100)$ based on the Fisher information matrix, bearing significant implications for understanding the formation mechanisms of double NS systems.

\end{abstract}

\maketitle

\section{Introduction}

Spinning neutron stars (NSs) are fascinating astrophysical objects that can emit hectohertz continuous gravitational waves (GWs), providing crucial insights into the state of dense matter \cite{Sieniawska2019, Riles2017, Lasky2015, 2022arXiv220606447R, Wette2023}. If such signals are detected, they can probe many aspects of dense matter physics  \cite{Pitkin2011, Soldateschi2021, Lu2022, Yim:2023nda, Feng2024}. 

The current continuous waveform templates of NSs used in the LIGO/Virgo/KAGRA \cite{advancedLIGO2015, AdvancedVirgo2015, KAGRA} searches are predominantly based on isolated NSs \cite{LIGO2022isolatedO3, LIGOisolatedNS2021, Abbott2022Narrowband, Abbott2022CasA} and NSs in binary systems with long orbital periods, where interactions with the companion star are neglected \cite{LIGONSinBinary2021, Covas2020, Covas2019, Zhang2021, Leaci2015, Pagliaro:2023bvi, Covas:2024nzs}. At present, there is still no credible detection of continuous GWs of NSs. However, the upper limits on GW emission from various sources and sky regions are constrained \cite{Ming2024, Owen2024, LiuandZou2022, Rajbhandari2021, Dergachev2019, Zhu2016}.

Binary population synthesis simulations based on the classical isolated binary evolution predict that upcoming space-based GW detectors, such as LISA \cite{LISA2017}, TianQin \cite{TianQin2016}, and Taiji \cite{Taiji2017}, have a promising capability to detect binaries with orbital period $P_{\rm b} \sim 10~{\rm min}$ and eccentricity $e \sim 0.01$ \cite{Andrews2020, Miao:2021awa, Wagg2022, Feng2023a}. The next-generation ground-based GW detectors, such as Einstein Telescope \cite{ET2010, Kalogera:2021bya} and Cosmic Explorer \cite{CosmicExplorer2022}, which will operate almost concurrently with LISA, TianQin, and Taiji, are very likely to detect spinning NSs in these binary systems. The continuous waveforms of the spinning NS components in these host binaries are shaped by various factors, one of which is the spin-orbit coupling \cite{poisson2014gravity}. Binaries with spins not aligned with the orbital angular momentum will precess around the direction of the total angular momentum, thereby introducing complexities in waveform patterns \cite{Feng2023b}. 
Accurate modeling of these patterns is crucial for detecting and characterizing the GWs emitted by such systems \cite{Feng2024}.
Because of their low orbital eccentricities, the waveform modeling of spinning NSs in double NS (DNS) systems formed in isolation within the Galactic field, as per Ref.~\cite{Feng2023b}, has been calculated for circular orbits during spin precession.

However, DNSs formed in dense stellar environments like globular clusters can exhibit larger eccentricities ($e>0.3$) compared to those formed in the Galactic field \cite{Kremer2018}. Additionally, a fast-merging DNS formation channel proposed to explain the presence of $r$-process elements in the Universe may also result in moderate to high eccentricities \cite{Andrews2020}.
In this paper, we generalize the waveform modeling from the spinning NSs in circular orbits to the case of eccentric orbits. Eccentricity influences from two aspects: its effect on the precession angular frequency of the NS's spin angular momentum around the total angular momentum of the binary system, and its effect on the Doppler frequency modulation of the NS as it orbits the binary barycenter (BB).
For NSs in binary systems, unlike isolated ones, we construct an inertial frame aligned with the binary's total angular momentum \cite{Feng2023b}. We calculate the rotation matrix between the body frame and this new inertial frame. By employing the perturbation method, we solve the precession equations arising from spin-orbit coupling to determine the spin precession angular frequency. We calculate the continuous waveform emitted by the spinning NSs using the quadrupole moment formula \cite{Einstein:1918btx}. These waveforms are systematically compared with those from circular orbits to assess the impact of orbital eccentricity. Finally, we estimate the parameter accuracy of these newly developed waveforms using the Cosmic Explorer as an illustration.

The paper is structured as follows. In Sec.~\ref{sec:NSwaveformunderSO}, we present the GW waveforms emitted by a NS with spin precession, incorporating the effects of orbital eccentricity. In Sec.~\ref{sec:comparisonwithcircular}, we compare these waveforms with those from NSs in circular binary systems. In Sec.~\ref{sec:detection}, by incorporating the Doppler modulation introduced by the eccentric orbital motion, we calculate the signal-to-noise ratio (SNR) and parameter estimation accuracy for NS waveforms in different DNS models with the Cosmic Explorer. Our conclusions are given in Sec.~\ref{sec:conclusion}.

\section{Gravitational waveforms from a triaxial NS under spin precession}
\label{sec:NSwaveformunderSO}

Consider a DNS system with component masses $m_{1,2}$, and an eccentric orbit characterized by a semimajor axis $a$ and eccentricity $e$. For the simple precession scenario with secular evolution \cite{Apostolatos1994, Barker1975}, where only one component NS of the DNS has spin angular momentum $\boldsymbol{S}$ and the spin-spin interaction is neglected \cite{poisson2014gravity, Feng2023b}, the precession equation for $\boldsymbol{S}$ takes a simplified form after averaging over one eccentric orbit:
\begin{align}\label{eq_dSdt}
\frac{\mathrm{d} \boldsymbol{S}}{\mathrm{d} t}= \boldsymbol{\Omega}_{\rm pre,av} \times \boldsymbol{S} \,.
\end{align}
This equation describes the motion of $\boldsymbol{S}$ around the binary's total angular momentum $\boldsymbol{J}$ at an angular velocity
\begin{equation}
\boldsymbol{\Omega}_{\rm pre,av} = \frac{G \boldsymbol{J}}{c^{2} a^{3}(1-e^2)^{3/2}}\left(2+\frac{3 m_2}{2 m_{1}}\right)  \,.
\end{equation}

\begin{figure}[t]
\centering
\includegraphics[scale=1.2]{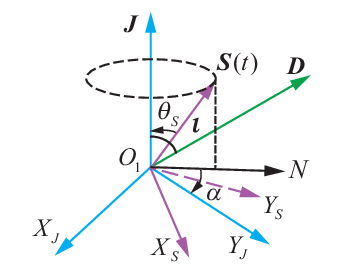}
\caption{The geometry of the $J$- and $S$-aligned frames. Both are fixed to the spinning NS with $O_1$ as the coordinate origin \cite{Feng2023b}. The opening angle for the precession cone of $\boldsymbol{S}$ is denoted by $\theta_S$. The $X_S$ axis is set to initially coincide with the $X_J$ axis, and $\boldsymbol{S}(t=0)$ is within the $Y_J-Z_J$ plane. The projection of $\boldsymbol{S}(t)$ onto the $X_J-Y_J$ plane is along ${O_{1}N}$, and it precesses with an angle $\alpha(t)$, given in Eq.~(\ref{eq_alpha}). The distant observer, represented by $\boldsymbol{D}$, is positioned at colatitude $\iota$ from $\boldsymbol J$ within the $Y_J-Z_J$ plane. }
\label{fig:JS_coord}
\end{figure}

Figure~\ref{fig:JS_coord} illustrates that the $J$-aligned inertial frame with coordinate system $(X_J, Y_J, Z_J)$ is centered at the center of mass ($O_1$) of the spinning NS, with the $Z_J$ axis oriented along the vector $\boldsymbol J$. In this frame, a distant observer lies within the $Y_J-Z_J$ plane at an inclination angle $\iota$, denoted by the position vector $\boldsymbol D$. The precession cone of the spin vector $\boldsymbol{S}$ has an opening angle $\theta_S$. The $S$-aligned frame $(X_S, Y_S, Z_S)$ has $Z_S$ directed along $\boldsymbol{S}$. We assume that initially the $X_S$ axis coincides with the $X_J$ axis and $\boldsymbol{S}$ is positioned in the $Y_J-Z_J$ plane. The evolution of $\boldsymbol{S}(t)$ is characterized by the precession angle 
\begin{equation}\label{eq_alpha}
\alpha(t) = \Omega_{\rm pre,av} t \,, 
\end{equation}
which is defined within the $X_J-Y_J$ plane. Here, ${\Omega}_{\rm pre,av}$ represents the magnitude of $\boldsymbol{\Omega}_{\rm pre,av}$.

For a freely rotating isolated rigid body, as described in Refs.~\cite{LANDAU1976, Zimmermann1980, Gao2020}, the $S$-aligned frame is typically used as the inertial frame. However, when analyzing spin precession, the $J$-aligned frame serves as the inertial frame due to angular momentum conservation.

%

Under the transverse-traceless gauge, the metric perturbation can be expressed as a combination of two GW polarizations, given by $h_{j k}^{\mathrm{TT}}=h_{+}\left(\hat{e}_{+}\right)_{j k}+h_{\times}\left(\hat{e}_{\times}\right)_{j k}$. Here, the polarization tensors are
$\hat{e}_{+} \equiv \hat{v} \otimes \hat{v}-\hat{w} \otimes \hat{w}  ,\, 
\hat{e}_{\times} \equiv \hat{v} \otimes \hat{w}+\hat{w} \otimes \hat{v} $,
with $\hat{v} \equiv \hat{e}_{y} \cos \iota-\hat{e}_{z} \sin \iota$ and $  \hat{w} \equiv -\hat{e}_{x} $ forming the orthogonal basis vectors transverse to the direction of wave propagation. The set \((\hat{e}_{x}, \hat{e}_{y}, \hat{e}_{z})\) serves as the basis vectors for the $J$-aligned frame.
The formulations for the two GW polarizations from the spinning NS are elaborated in Ref.~\cite{Zimmermann1980},
\begin{subequations}
\label{eq:waveform_h}
\begin{align}\label{eq:waveform_hplusRA}
h_{+} =& -\frac{G}{c^4 D} [({\mathcal R}_{y\mu} \cos \iota - {\mathcal R}_{z\mu} \sin \iota) \\ \nonumber
&\times ({\mathcal R}_{y\nu} \cos \iota - {\mathcal R}_{z\nu} \sin \iota)-{\mathcal R}_{x\mu} {\mathcal R}_{x\nu}] {\mathcal A}_{\mu \nu} \,, \\  \label{eq:waveform_hcrossRA}
h_{\times} =&\, \frac{2 G}{c^4 D} ({\mathcal R}_{y\mu} \cos \iota - {\mathcal R}_{z\mu} \sin \iota) {\mathcal R}_{x\nu} {\mathcal A}_{\mu \nu}  \,,
\end{align}
\end{subequations}
where Einstein summation is applied to the indices $\mu$ and $\nu$, each ranging over the set $\{1,2,3\}$. 
The matrix $\mathcal{R}$ denotes the rotation matrix that transforms from the body frame (constructed by three principal axes of inertia) of the spinning NS to the $J$-aligned inertial frame. This matrix is achieved through the following rotation transformations: initially from the body frame to the $S$-aligned frame using the matrix $R$ [see Eq. (24) in Ref.~\cite{Zimmermann1980}], followed by a transformation from the $S$-aligned frame to the $J$-aligned frame using $\mathcal{T}_{S\rightarrow J}$. Therefore,
\begin{equation}\label{eq_newR}
\mathcal{R} = \mathcal{T}_{S\rightarrow J} \cdot R \,,
\end{equation}
where
\begin{align}\label{RbodytoJ}
\mathcal{T}_{S\rightarrow J} = 
\left(
\begin{array}{ccc}
 \cos \alpha  & - \cos \theta_S \sin \alpha  & -\sin \theta_S \sin \alpha  \\
 \sin \alpha  & \cos \theta_S \cos \alpha  & \sin \theta_S \cos \alpha \\
 0 & -\sin \theta_S & \cos \theta_S \\
\end{array}
\right) \,.
\end{align}
%

The symmetric matrix $\mathcal{A}$ in Eqs.~(\ref{eq:waveform_h}) is defined in terms of three principal moments of inertia ($I_1, I_2, I_3$) and their corresponding angular velocities ($\omega_1,\omega_2,\omega_3$) \cite{Zimmermann1980},
\begin{subequations}
\label{eq_definemathcalA}
\begin{align}
\mathcal{A}_{11} &= 2\left(\Delta_{2} \omega_{2}^{2}-\Delta_{3} \omega_{3}^{2}\right) \,,\\ 
\mathcal{A}_{12}  &= \left(\Delta_{1}-\Delta_{2}\right) \omega_{1} \omega_{2}+\Delta_{3} \dot{\omega}_{3}  \,,
\end{align}
\end{subequations}
with $\Delta_{1} \equiv I_{2}-I_{3}$, $\Delta_{2} \equiv I_{3}-I_{1}$, $\Delta_{3} \equiv I_{1}-I_{2}$. 
The other components follow by symmetry and cycling the indices $1\rightarrow2\rightarrow3\rightarrow1$.

The gravitational waveforms emitted by the spinning NS under spin precession are closely related to the spin angular velocity, making it crucial to solve Eq.~(\ref{eq_dSdt}). 
Assume that the spinning NS is a rigid body, with a spin period of $P_{\rm s}$, and the magnetic dipole moment is $p$ (in the Gaussian centimeter-gram-second unit).
Other possible sources of external torque that can be added to the right-hand side of Eq.~(\ref{eq_dSdt}) include gravitational radiation and magnetic dipole radiation, whose magnitudes are given by the following expressions \cite{Ostriker1969, Gao:2022hzd}:
\begin{subequations}\label{eq_othertorque}
\begin{align} 
\left(\frac{\mathrm{d} S}{\mathrm{d} t}\right)_{\mathrm{gr}} &\sim -\frac{1024 G \pi^5}{5 c^5} \frac{(I_1 - I_2)^2}{P_{\rm s}^5} \,, \\    
\left(\frac{\mathrm{d} S}{\mathrm{d} t}\right)_{\mathrm{md}} &\sim -\frac{16 \pi^3}{3 c^3 } \frac{p^2 }{P_{\rm s}^3}  \,.
\end{align}
\end{subequations}

For a spinning NS with a magnetic field of $10^9~\rm{G}$ and a spin period of $10~\rm{ms}$ in a tight DNS with a 10-min orbital period (see Table~\ref{tab:parameters}), the rates for the change of angular velocity are
\begin{subequations}\label{eq_dif_domegadt}
\begin{align} 
\left(\frac{\mathrm{d} \omega}{\mathrm{d} t}\right)_{\mathrm{so}} &\sim 2\times 10^{-4} ~\rm{Hz/s} \,, \\ 
\left(\frac{\mathrm{d} \omega}{\mathrm{d} t}\right)_{\mathrm{gr}} &\sim -3\times10^{-16} ~\rm{Hz/s} \,, \\    
\left(\frac{\mathrm{d} \omega}{\mathrm{d} t}\right)_{\mathrm{md}} &\sim -1\times10^{-16}  ~\rm{Hz/s} \,.
\end{align}
\end{subequations}
Hence, for simplicity, we ignore the effects of gravitational radiation and magnetic dipole radiation on $\boldsymbol{S}$ in the following analysis.

Waveforms from a triaxial NS are typically represented through a series expansion using certain small parameters, as detailed in Refs.~\cite{Zimmermann1980, Broeck2005, Gao2020, Feng2023b}.
To simplify the calculations, we specify the angular frequencies of free precession $\Omega_{\rm p}$ and rotation $\Omega_{\rm r}$ for the spinning NS,
along with three small parameters (oblateness or poloidal ellipticity $\epsilon$, nonsphericity $\kappa$, and  wobble angle $\gamma$) delineating the NS's characteristics 
\begin{align}\label{eq_smallparas}
\epsilon \equiv \frac{I_3-I_1}{I_3} \,,\quad
\kappa   \equiv \frac{1}{16}\frac{I_3}{I_1}\frac{I_2-I_1}{I_3-I_2} \,,\quad
\gamma   \equiv \frac{aI_1}{bI_3} \,,
\end{align}
where $a$ and $b$ are the initial angular velocities for $\omega_1$ and $\omega_3$. The typical values for these parameters, discussed in Ref.~\cite{Broeck2005}, are $\epsilon \ll \kappa \ll \gamma_{\rm max}$, with $\kappa \sim O(\gamma^2)$. We aim to expand the waveforms to order $O(\gamma^2)$ and $O(\kappa)$.
As in Ref.~\cite{Feng2023b}, we use the perturbation method to solve Eq.~(\ref{eq_dSdt}) analytically in the body frame, $\boldsymbol{S}=S_{\mu}\boldsymbol{e}_{\mu}$ 
and 
$\boldsymbol{\Omega}_{\rm pre} = {\Omega}_{\rm pre} \mathcal{R}_{z\mu} \boldsymbol{e}_{\mu}$, by assuming $\omega_{i} \approx \Omega_{i} + \delta \Omega_{i} \quad (i=1,2,3)$.
Finally, the waveforms produced by the spinning NS experiencing spin precession can be expanded as follows
\begin{subequations}  \label{eq:waveform:components}
\begin{align}
h_+ &= h_+^{(1)} + h_+^{(2)} + h_+^{(3)} + \cdots \\
h_\times &= h_\times^{(1)} + h_\times^{(2)} +  h_\times^{(3)} + \cdots
\end{align}
\end{subequations}
where
\begin{widetext}
\begin{align} 
\nonumber
 h_{+}^{(1)}= &\frac{G b^2 I_3 \epsilon \gamma}{4 c^4 D}\bigg\{ \Big[\sin 2 \theta_S \big(6 \sin ^2 \iota-(3+\cos 2 \iota) \cos 2 \alpha \big)+4 \cos 2 \theta_S \cos \alpha \sin 2 \iota \Big]  \cos \left[\left(\Omega_{\mathrm{p}}+\Omega_{\mathrm{r}}\right) t\right] \\ 
&  +2\Big[-2 \cos \theta_S \sin 2 \iota \sin \alpha+ \big(3+\cos 2 \iota \big) \sin \theta_S \sin 2 \alpha \Big] \sin \left[\left(\Omega_{\mathrm{p}}+\Omega_{\mathrm{r}}\right)t\right]\bigg\} , \\ \nonumber
\end{align}
\end{widetext}
and the rest of the first three waveform components can be found in Ref.~\cite{Feng2023b} by substituting $\Omega_{\rm pre}t$ with $\alpha$ [cf. Eq.~(\ref{eq_alpha})]. 
In the results for secular evolution of spin precession, the waveforms emitted in the NS frame are mainly influenced by orbital eccentricity through the precession angular frequency. The Doppler shift due to the NS's eccentric orbit around the BB will be discussed in Sec.~\ref{sec:detection}.

\begin{figure*}[t]
\centering
\includegraphics[scale=0.7]{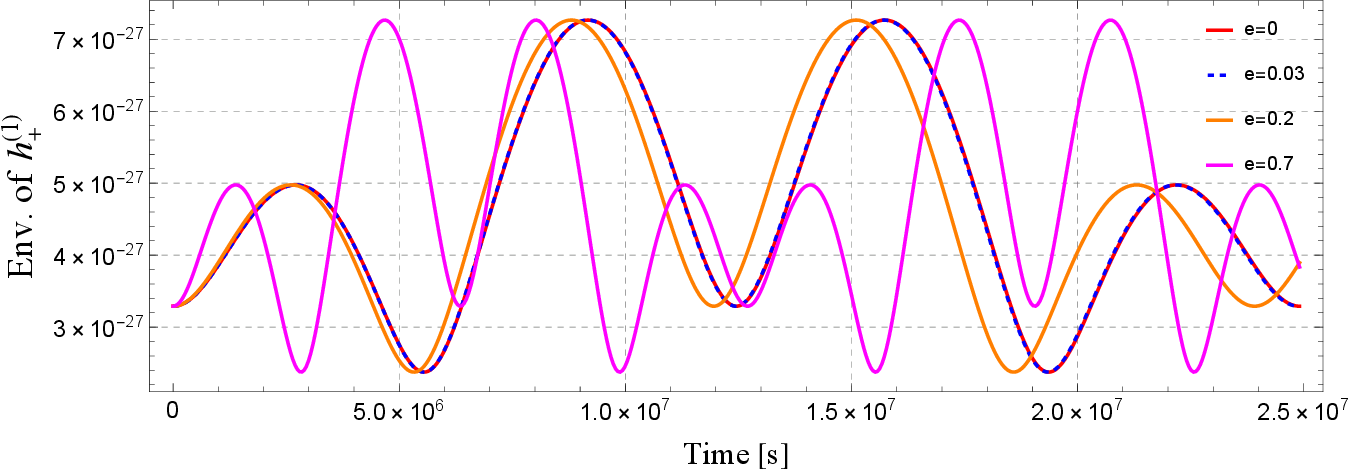}
\caption{The amplitude envelope (taking only positive values) of $h_{+}^{(1)}$ for different eccentricities, assuming a binary orbital period of 10 min, with the NS parameters listed in Table~\ref{tab:parameters}. 
}
\label{fig:envwaveform}
\end{figure*}
\begin{table}[t]
    \centering
    \caption{Parameters and their values for the NS example and its geometry in the host binary. }
    \label{tab:parameters}
    \renewcommand{\arraystretch}{1.4}
    \begin{ruledtabular}
    \begin{tabular}{lcc}
    Parameter & Symbol & Value  \\ \hline
    Spin period & $P_{\rm s}$ & $10~{\rm ms}$  \\
    Moment of inertia & $I_3$ & $2.0\times 10^{38}~{\rm kg~m^2}$  \\
    Oblateness/poloidal ellipticity & $\epsilon$ & $3.6\times 10^{-6}$  \\
    Nonsphericity  & $\kappa$ & $1.75\times 10^{-4}$  \\
    Wobble angle & $\gamma$ & $5.0\times 10^{-2}$  \\
    Equatorial ellipticity & $\varepsilon$ & $1.0\times 10^{-8}$  \\
    Initial frequency for $\omega_3$ & $b$ & $627.53~{\rm Hz}$  \\
    Rotation frequency & $\Omega_{\rm r}$ & $628.32~{\rm Hz}$  \\
    Free precession frequency & $\Omega_{\rm p}$ & $2.26\times 10^{-3}~{\rm Hz}$  \\
    Binary inclination & $\iota$ & $\pi/4$ \\
    Precession cone opening angle & $\theta_S$ & $5\pi/12$ \\
    Distance & $D$  & $1~{\rm kpc}$  \\
    \end{tabular}
    \end{ruledtabular}
\end{table}

Figure~\ref{fig:envwaveform} illustrates the positive amplitude envelope of $h_{+}^{(1)}$ for various eccentricities, assuming a binary orbital period of 10 minutes, and using the NS parameters listed in Table~\ref{tab:parameters}.

\section{Comparison with waveforms from NS in circular-orbit case}
\label{sec:comparisonwithcircular}

We can use the fitting factor (FF) to assess the match between two types of waveforms quantitatively. The FF between the actual GW waveforms produced by a spinning NS in an eccentric binary ($h_{\rm e}$) and those in a circular one ($h_{\rm c}$) is defined as \cite{Apostolatos1995FF}:
\begin{equation}
{\rm{FF}} \equiv \max _{\boldsymbol {\lambda}} \frac{\left(h_{\rm e}, h_{\rm c}\right)}{\sqrt{\left(h_{\rm e}, h_{\rm e}\right)(h_{\rm c}, h_{\rm c})}} \,,
\end{equation}
where ${\boldsymbol{\lambda}}$ represents the set of parameters characterizing the waveforms. For a quasimonochromatic signal, the inner product $\left(h_{\rm e}, h_{\rm c}\right)$ simplifies to $\left(h_{\rm e}, h_{\rm c}\right) \equiv \int_{0}^{T_{\rm obs}} h_{\rm e}(t) h_{\rm c}(t) \mathrm{d} t$, where $T_{\rm obs}$ denotes the observation time.
We assume that the host binary, where the spinning NS is located, can be detected by the space-based GW detectors (e.g., LISA, TianQin, and Taiji).
We choose a threshold of 0.97 (0.9) for FF, corresponding to a $\sim 10\%$ ($27\%$) reduction in the detection rate \cite{Apostolatos1995FF, Mukherjee:2022tuc}. 

%
\begin{figure*}[t]
\centering
\includegraphics[scale=0.68]{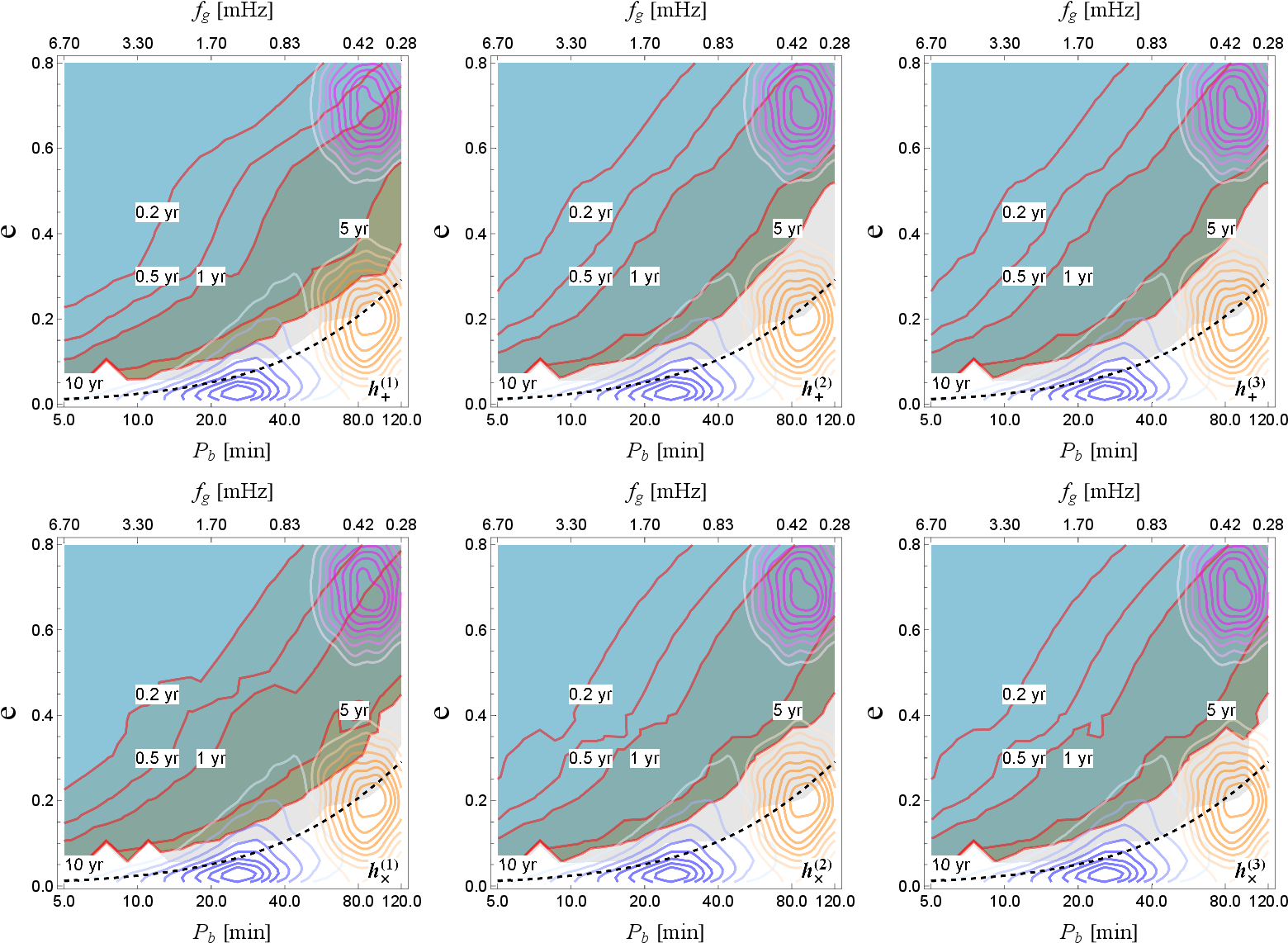}
\caption{The time required for the FF to decrease to 0.97 for the first three NS waveform components in binaries [see Eqs.~(\ref{eq:waveform:components})]. The parameters used here for the spinning NS can be found in Table~\ref{tab:parameters}. The red solid lines in each subplot correspond to contours for 0.2, 0.5, 1, 5, and 10 yr. The abscissa and ordinate are the orbital period ($P_{\rm b}$) and eccentricity ($e$) of the binary respectively, and the top abscissa is the second harmonic frequency of orbit ($f_{\rm g}=2/P_{\rm b}$). The ranges of $P_{\rm b} \in [5,120]~{\rm min}$ and $e \in [0.01,0.8]$ fall in the millihertz band of the space-borne GW detectors. The blue contours represent the DNS population from COMPAS simulation through isolated binary evolution \cite{Wagg2022}, while the orange and magenta contours denote two putative fast-merging formation channels, i.e., fast-merging isolated binaries and dynamically formed binaries, respectively \cite{Andrews2020}. These three types of contours indicate the percentage of the enclosed DNS population, from inner to outer, ranging from $10\%$ to $80\%$ in $10\%$ increments. The black dashed line in the subplot illustrates the evolution track of the Hulse-Taylor binary (PSR~B1913+16), assuming that orbital decay results exclusively from gravitational radiation.
For a tight binary system with a 10-min orbital period and an eccentricity below 0.1, the waveforms emitted by the component NS in an eccentric binary remain closely aligned with those from a circular orbit for 5 yr of observation.
}
\label{fig:timecontour_ecc_pb}
\end{figure*}

Figure \ref{fig:timecontour_ecc_pb} illustrates the time it takes for the FF to decrease to 0.97 for the first three NS waveform components in Eqs.~(\ref{eq:waveform:components}). 
The horizontal and vertical axes represent the orbital periods ($P_{\rm b}$) and eccentricities ($e$) of the binaries, respectively. The secondary horizontal axis on top denotes the second harmonic frequency of orbital motion ($f_{\rm g} = 2/P_{\rm b}$). These panels present DNS populations with equal component masses, $m_1=m_2=1.4M_{\odot}$. The orbital period of these binaries ranges from 5 min to 2 h, covering the frequency band of space-based GW detectors (e.g., LISA, TianQin, and Taiji), with eccentricities taken from 0.01 up to 0.8.  
The characteristic parameters of the spinning NS are given in Table~\ref{tab:parameters} \cite{Feng2023b}.
In each subplot, the red solid lines indicate contours for 0.2, 0.5, 1, 5, and 10 yr. 
As seen in this figure, in the case of a tight binary system with a 10-min orbital period and an eccentricity less than 0.1, the waveforms emitted by the NS components remain closely aligned for 5 yr, with no significant difference from the circular orbit.
The subfigures also depict the $P_{\rm b}-e$ evolution track of the Hulse-Taylor binary (PSR~B1913+16) \cite{HulseTaylor1975} within the parameter space as a black dashed line. The results indicate that we do not need to consider the impact of residual eccentricity on the GW radiation waveform of NSs in Hulse-Taylor-like binaries when they enter the relevant frequency band (see e.g., Ref.~\cite{Zhao:2021bjw}). 

Figure \ref{fig:timecontour_ecc_pb} also depicts the parameter distributions in $P_b-e$ space provided by three DNS population models. These three types of contours indicate the percentage of the enclosed DNS population, from inner to outer, ranging from $10\%$ to $80\%$ in $10\%$ increments. The first model, detectable by LISA and represented by the blue contours, originates from Ref.~\cite{Wagg2022}. This model  (abbreviated as COMPAS model) simulates compact binary populations in the Milky Way using the rapid population synthesis code COMPAS \cite{COMPAS}, based on the classical isolated binary evolution channel. The simulation incorporates a novel, empirically based analytical model for the metallicity-dependent star formation history of our Galaxy. Although most of the DNSs in the simulation have an equal mass of approximately $1.3~M_\odot$, it is expected that this should not significantly deviate from the $1.4~M_\odot$ case that we adopt in this study. For this DNS population model, we find that for over $80\%$ of the sources (where $e<0.3$), a noticeable deviation from the circular-orbit case would only occur after more than 10 yr of observation.
The second model (denoted by orange contours) and the third model (denoted by magenta contours) come from the putative fast-merging DNS evolutionary channel \cite{Andrews2020}. For these two models, DNSs are randomly created by using a log-normal distribution for orbital separation with parameters $\mu = -0.1$ and $\sigma = 0.2$ in units of solar radius. The eccentricity follows a normal distribution with means of $\mu = 0.2$ and $\mu = 0.7$ for the two models, both having a standard deviation of $\sigma = 0.1$ and truncated to ensure that $0 < e < 1$. For the second model with fast-merging isolated binaries (abbreviated as IsolatedFast model), the majority of the DNSs with eccentricities less than 0.3 do not exhibit deviations from the circular orbit within 10 yr of observation. For the third model, which involves fast-merging dynamically formed binaries (abbreviated as DynamicalFast model), most sources will exhibit deviations from the circular-orbit case within 0.5-2 yr for $h_{+,\times}^{(1)}$. For $h_{+,\times}^{(2)}$ and $h_{+,\times}^{(3)}$, deviations are expected within 1-5 yr.
For comparison, we have calculated the time required for FF to decline to 0.9 for various NS waveform components. In the DynamicalFast model, the majority of sources will display deviations from the circular-orbit case within 1-5 yr for $h_{+,\times}^{(1)}$. For $h_{+,\times}^{(2)}$ and $h_{+,\times}^{(3)}$, deviations are anticipated within 2-10 yr.

Additionally, we found that the time required for the FF to drop to 0.97 ($T_{97}$) is related to the spin-induced precession period ($T_{\rm pre} = 2\pi/\Omega_{\rm pre}$) by the ratio 
\begin{equation}
T_{97}/T_{\rm pre} \approx 0.04/e^2 \,. 
\end{equation}
This ratio is independent of the orbital period and inversely proportional to the square of the eccentricity. The results indicate that when $e \approx 0.2$, $T_{97} \approx T_{\rm pre}$; as $e \to 1$, $T_{97}/T_{\rm pre} \to 0.04$; and as $e \to 0$, $T_{97}/T_{\rm pre} \to \infty$.

\section{Detecting GW from a spinning NS in a binary}
\label{sec:detection}

\subsection{Signal model in detector frame}
\label{subsec_signalmodel}

The waveforms for the spinning NS in a binary system are obtained by incorporating the Doppler frequency shifts that arise from the NS's orbital motion around the BB into the waveform phases. The signal detected can be obtained by including the antenna pattern function of the detector and its Doppler modulation relative to the Solar System barycenter (SSB) \cite{LIGONSinBinary2021}.

First, we incorporate Doppler modulation into the waveform phases from the spinning NS. Refer to the conventions in Ref.~\cite{Feng2023b}, although in this work an eccentric orbit is considered.
The position vector $\boldsymbol{r}_{1}$ of the spinning NS in the binary coordinate system is given by 
\begin{align}\label{rL2rJ}
\boldsymbol{r}_{1} = r_1 \left(
\begin{array}{c}
\cos (\omega_{\rm b}t) \cos \alpha - \cos \theta _L \sin (\omega_{\rm b}t) \sin \alpha  \\
\cos \theta _L \sin (\omega_{\rm b}t) \cos \alpha + \cos (\omega_{\rm b}t) \sin \alpha  \\
\sin \theta _L \sin (\omega_{\rm b}t)  \\
\end{array}
\right) \,, 
\end{align}
with the orbital angular frequency $\omega_{\rm b} = 2\pi /P_{\rm b}$.
The magnitude of $\boldsymbol{r}_{1}$ is 
\begin{equation}
r_1 = \frac{a_1 (1-e^2)}{1+e \cos{f_{\rm ta}}} \,,
\end{equation}
where $f_{\rm ta}$ is the true anomaly measured from the periapsis to the direction of the orbital separation, and the semimajor axis of the NS orbit is $a_1=G^{1/3}m_2/(\omega_{\rm b}M)^{2/3}$.
The true anomaly $f_{\rm ta}$ can be expanded in terms of the mean anomaly $M_{\rm A} = \omega_{\rm b}(t - t_0)$ as follows,
\begin{align}
\cos{f_{\rm ta}} = -e + \frac{1-e^2}{e} \sum_{n=1}^{\infty} 2 J_n(ne) \cos\left({nM_{\rm A}}\right) \,.
\end{align}

In the detector coordinate system, the Doppler phase shift of the GW angular frequency $\Omega$ from the spinning NS in a binary can be written as the combination of detector Doppler shift relative to the SSB and source Doppler shift relative to the BB, 
\begin{equation}\label{eq:Dopshift}
\Delta \Omega_{\rm{D}} \approx  \frac{\Omega}{c} \left({\boldsymbol n}  \cdot \frac{{\rm d}\boldsymbol{r}_{\rm d}}{{\rm d}t}+{\boldsymbol n}_{\rm b}\cdot \frac{ {\rm d}\boldsymbol{r}_1}{{\rm d}t}\right) \,,
\end{equation}
with $\boldsymbol{r}_{\rm d}$ the detector's position vector and  $\boldsymbol{n}$ the spinning NS's line of sight in the SSB coordinate system \cite{Jaranowski1998}. Therefore,
$-{\boldsymbol n}_{\rm b}=(0,\sin{\iota},\cos{\iota})$ is the position of the SSB in the binary coordinate system.
The Doppler-modulated waveforms $H_{+,\times}^{(n)}(t)$ can be obtained through $h_{+,\times}^{(n)}$ by replacing $\Omega \rightarrow \Omega + \Delta \Omega_{\rm{D}}$, with $\Omega \in \{\Omega_{\rm r}, \Omega_{\rm p}, \Omega_{\rm pre,av} \}$.

Then GW strain signal from a spinning triaxial NS in a binary can be expressed as a sum of different waveform components $h_n(t)~(n=1,2,3)$ as follows \cite{Feng2023b}
\begin{align}\label{eq:ht}
h(t) = \sum_{n=1}^3 h_n(t) = \sum_{n=1}^3 F_{+}(t) H_{+}^{(n)}(t) + F_{\times}(t) H_{\times}^{(n)}(t)   \,,
\end{align}
where $F_{+,\times}(t)$ are the antenna pattern functions of the GW detector (see Ref.~\cite{Jaranowski1998} for the explicit expressions). 
The functions $F_{+,\times}(t)$ depend on the following parameters: $\gamma_{\rm o}$, which defines the detector's orientation to the local geographical directions; $\zeta$, the angle between the arms of the interferometer; $\lambda$, the geographical latitude of the detector location; $(\alpha,\delta)$, the right ascension and declination of the source; $\psi_{\rm p}$, the GW polarization angle; $\Omega_{\rm Er}$, Earth's rotational angular frequency; and $\phi_{\rm r}$, the initial phase of Earth's diurnal rotation.

\subsection{Effects of eccentricity on parameter estimation}

We utilize the Fisher information matrix (FIM) to provide a quantitative evaluation of parameter accuracy for GW detection, as given in e.g. Ref.~\cite{Jaranowski1999PhRvD}. For a GW signal $h(t)$ with parameter set $\boldsymbol{\lambda}$, the FIM is defined as:
\begin{equation}\label{eq_FIM}
{{\Gamma }^{ij}} \equiv \left( \frac{\partial h}{\partial {{\lambda }_{i}}}, \frac{\partial h}{\partial {{\lambda }_{j}}} \right).
\end{equation}
For a monochromatic signal of frequency $f$, the noise-weighted inner product is simplified and approximated as $\left(a, b\right) \simeq \frac{2}{S_{n}\left(f\right)} \int_{0}^{T_{\rm obs}} a(t) b(t) dt$ \cite{Shah2012}, where $S_{n}(f)$ represents the power spectral density of instrumental noise at frequency $f$, and $T_{\rm obs}$ is the observation time. The optimal  SNR for GW detection is defined as ${\rm SNR} \equiv (h,h)^{1/2}$. Three different waveform components correspond to different SNRs (${\rm SNR}_1, {\rm SNR}_2, {\rm SNR}_3$) and FIMs ($\Gamma_1, \Gamma_2, \Gamma_3$), with the total FIM given by $\Gamma = \Gamma_1 + \Gamma_2 + \Gamma_3$. The root-mean-square (rms) error of parameter ${\lambda_i}$ is estimated as
\begin{equation}
\Delta {\lambda_i} = \sqrt{{\Sigma_{ii}}} \,, 
\end{equation}
where the covariance matrix $\Sigma$ is the inverse of the FIM $\Gamma$, i.e., $\Sigma = {\Gamma^{-1}}$.
The amplitudes of different waveform components are defined as $h_{10}={{2G{b^2}{I_3}\epsilon \gamma }}/({c^4}D)$, $h_{20}={{64G{b^2}{I_3}\epsilon \kappa }}/({c^4}D)$, $h_{30}={4Gb^2{I_3}\epsilon \gamma^2}/({c^4}D)$. The parameter set in this work is $\boldsymbol{\lambda}= \big\{ \ln{h_{10}}, \ln{h_{20}}, \ln{h_{30}}, \alpha, \sin{\delta}, \ln{P_{\rm b}}, \ln{e}, \cos{\iota}, \theta_S, \psi_{\rm p}, \ln{\Omega_{\rm r}},\\ \ln{\Omega_{\rm p}} \big\}$. Logarithmic scales are used for some parameters because relative errors offer more meaningful insights than absolute errors in these parameters, e.g., $\Delta \ln{h_{10}}=\Delta h_{10}/h_{10}$ represents the relative error in $h_{10}$.
Following the sky localization error defined in Ref.~\citep{cutler1998lisa},  for a source at $(\alpha,\delta)$ we have $\Delta \Omega = 2 \pi \big(\Sigma_{\alpha \alpha} \Sigma_{\sin \delta \sin \delta } -\Sigma_{\alpha \sin \delta }^{2} \big)^{1/2}$.

\begin{figure*}[t]
\centering
\includegraphics[scale=0.65]{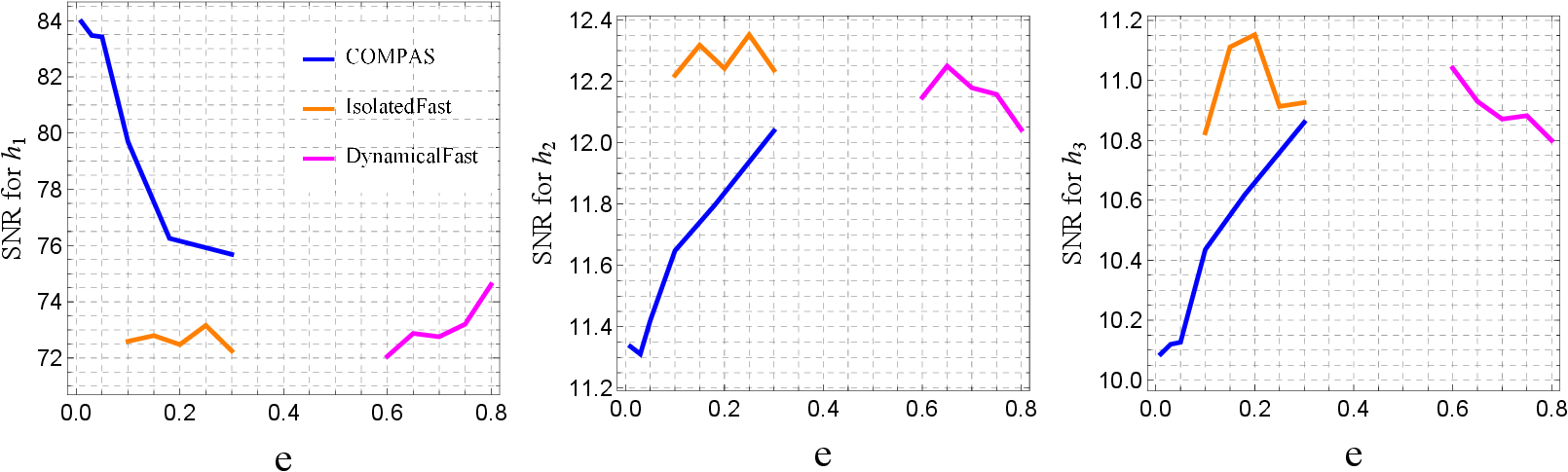}
\caption{The SNR as a function of eccentricity is estimated for the first three waveform components, $h_1(t), h_2(t), h_3(t)$, from a spinning NS in different DNS models using the Cosmic Explorer. The parameters used here can be found in Tables~\ref{tab:parameters} and \ref{tab:detectparameters}. }
\label{fig:snr_ecc}
\end{figure*}

We employ the Cosmic Explorer \cite{CosmicExplorer2022} as an illustration for the following FIM analysis. A 40 km arm length optimized for low frequencies is assumed, and the amplitude spectral density (ASD) of the instrumental noise is $\sqrt{S_{n}(f_0)} = 1.36\times 10^{-25}~{\rm Hz}^{-1/2}$ and $\sqrt{S_{n}(2f_0)} = 1.66\times 10^{-25}~{\rm Hz}^{-1/2}$ for $f_0=100~{\rm Hz}$, with an observation time of $T_{\rm obs}=0.5~{\rm yr}$. Angular parameters are chosen as $\zeta=\pi/2$, $\lambda=0.764$, $\gamma_{\rm o}=1.5$, $\phi_{\rm r}=\phi_{\rm o}=0$, $\alpha=1.209$, $\delta=1.475$, and $\psi_{\rm p}=1.0$. These main parameters are all summarized in Table~\ref{tab:detectparameters}.

\begin{table}[htb]
    \centering
    \caption{Parameters and their values for the DNS and the Cosmic Explorer detector. }
    \label{tab:detectparameters}
    \renewcommand{\arraystretch}{1.4}
    \begin{ruledtabular}
    \begin{tabular}{lcc}
    Parameter & Symbol & Value  \\ \hline
    Source sky position & $(\alpha,\delta)$  & $(1.209, 1.475)$  \\
    GW polarization angle & $\psi_{\rm p}$ & 1.0  \\
    Detector orientation & $\gamma_{\rm o}$   & 1.5  \\
    Detector geographical latitude & $\lambda$   & 0.764   \\
    Interferometer arm &  &  \\
    Opening angle & $\zeta$  & $\pi/2$  \\
    Noise ASD at $f_0=100~{\rm Hz}$ & $\sqrt{S_{n}(f_0)}$ & $1.36\times 10^{-25}/\sqrt{\rm Hz}$  \\
       & $\sqrt{S_{n}(2f_0)}$  & $1.66\times 10^{-25}/\sqrt{\rm Hz}$ \\
    Observation time & $T_{\rm obs}$  & 0.5~yr
    \end{tabular}
    \end{ruledtabular}
\end{table}

The FIM is numerically calculated using \textit{Mathematica} with parameters in Tables~\ref{tab:parameters} and \ref{tab:detectparameters}, for different binary parameters $(P_{\rm b}, e)$. For the COMPAS model, the orbital eccentricity is set to $e \in \big\{0.01, 0.03, 0.05, 0.1, 0.18, 0.30 \big\}$, and the most likely values of the orbital period distribution are $P_{\rm b} \in \big\{1535, 1535, 1535, 1838, 2200, 2634 \big\}~{\rm s}$. In terms of the IsolatedFast and DynamicalFast fast-merging DNS evolutionary channels, the eccentricities are set to $e \in \big\{0.1, 0.15, 0.2, 0.25, 0.3 \big\}$ and $e \in \big\{0.6, 0.65, 0.7, 0.75, 0.8 \big\}$, respectively, and the corresponding orbital period for both scenarios is assumed to be the mean value of the distribution, which is  $5155~{\rm s}$ (see Fig.~\ref{fig:timecontour_ecc_pb}).

\begin{figure*}[t]
\centering
\includegraphics[scale=0.65]{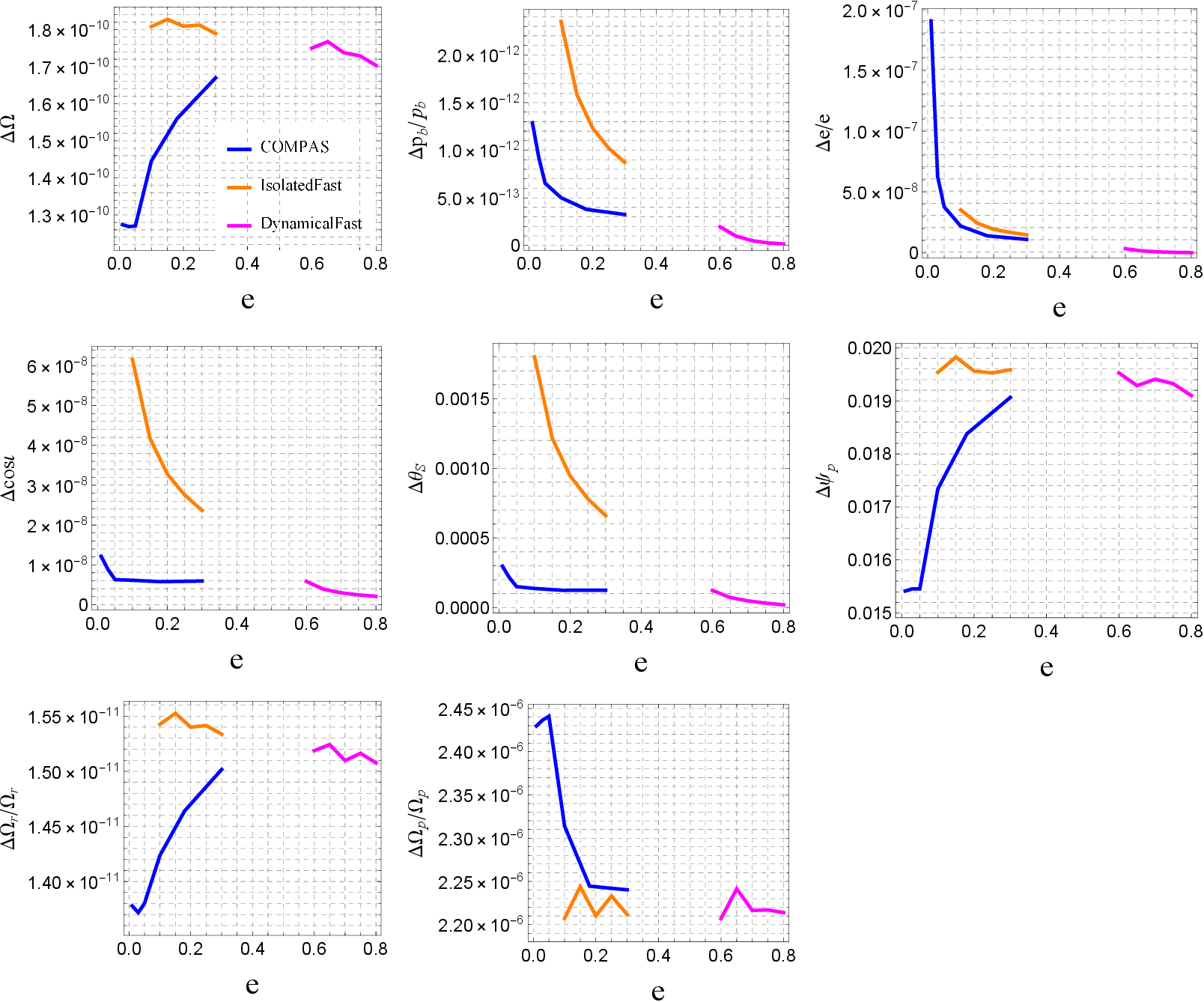}
\caption{The rms error as a function of eccentricity is estimated through the FIM method for the first three waveform components, $h_1(t), h_2(t), h_3(t)$, from a spinning NS in different DNS models using the Cosmic Explorer. The parameters used here are the same as those in Fig.~\ref{fig:snr_ecc}.}
\label{fig:RMSerror-Ecc}
\end{figure*}

The SNR as a function of eccentricity for the first three waveform components from a spinning NS in different binaries is shown in Fig.~\ref{fig:snr_ecc}, where, from left to right, subplots correspond to the waveform components $h_1(t)$, $h_2(t)$, and $h_3(t)$, respectively. In the figure, the blue, orange, and magenta colors represent the DNS population from the COMPAS, IsolatedFast, and DynamicalFast models, respectively. The three colors are the same as in Figs.~\ref{fig:timecontour_ecc_pb} and \ref{fig:RMSerror-Ecc} below, for ease of comparison. For the waveform component $h_1(t)$, the SNR of the NS waveforms in DNSs from the COMPAS model decreases as eccentricity increases, whereas in the IsolatedFast and DynamicalFast models, the SNR increases with increasing eccentricity. The results for the waveform components $h_2(t)$ and $h_3(t)$ are different.

Changes in SNR and its corresponding rms errors (as shown in Fig.~\ref{fig:RMSerror-Ecc}) can be understood through variations in the angular velocity of spin precession. For the COMPAS model, as eccentricity increases, the orbital period lengthens (see Fig.~\ref{fig:timecontour_ecc_pb} and six points chosen in the text), resulting in a decrease in precession angular velocity. This means the accumulated GW signal by spin precession decreases over the same observation time (see Fig.~\ref{fig:envwaveform}). Conversely, in the IsolatedFast and DynamicalFast models, with increasing eccentricity (while keeping the orbital period constant), the precession angular velocity rises, leading to more accumulated GW signal over the same period, thus increasing the SNR.

Figure~\ref{fig:RMSerror-Ecc} shows how the rms errors of the spinning NS waveform parameters vary with eccentricity in three different DNS models.
Since the relative errors for three amplitudes are inversely proportional to their SNRs, we do not show them in Fig.~\ref{fig:RMSerror-Ecc}.
In the case of DNSs within the IsolatedFast and DynamicalFast models, the accuracy of all parameters improves (i.e., the errors decrease) with increasing SNR. However, for the DNSs in the COMPAS model, parameters such as $\ln{P_{\rm b}}, \ln{e}, \cos \iota, \theta_S$  exhibit the opposite behavior. For this particular behavior, we conducted a test. By setting the orbital period as a constant in the COMPAS model and only varying the eccentricity, the behavior of the rms errors for these parameters resembles that observed in the other two models. This indicates that in the COMPAS model, the unusual behavior of those parameters is caused by the simultaneous variation of the orbital period and eccentricity where the orbital period increases with
increasing eccentricity (see Fig.~\ref{fig:timecontour_ecc_pb} and six points chosen in the text). An exception is the estimation of $\Omega_{\rm p}$, where the parameter accuracy decreases (i.e., the errors increase) as the SNR decreases.
The estimate for $\Delta e/e \sim 10^{-7}$ here is more precise than the eccentricity measurements ($\sim 10^{-2}$) from inspiraling GWs of the binary in the millihertz band \cite{Feng2023a}, which will have significant implications for understanding the formation mechanisms of DNSs.

\section{Conclusions}\label{sec:conclusion}

In this study, we calculate the continuous GW waveforms emitted by spinning NSs in binaries that experience spin precession due to spin-orbit coupling, incorporating the influence of orbital eccentricity. We compared these waveforms with those emitted by NSs in circular binary systems. By calculating the time required for the FF to decrease to 0.97 for the first three NS waveform components in binaries, we found that for DNS systems generated by the COMPAS and IsolatedFast models, where the orbital eccentricities are relatively small, there is no significant impact of orbital eccentricity on the waveforms emitted by spinning NSs in binaries, indicating the waveforms without eccentricity is sufficiently accurate in the detection of these sources. However, for DNS systems generated by the DynamicalFast model, where the orbital eccentricities are larger, the orbital eccentricity has a more significant impact on the waveforms. Additionally, by accounting for the Doppler modulation effect introduced by eccentric orbital motion, we assessed the SNR and the accuracy of parameter estimation for NS waveforms across various DNS models using the Cosmic Explorer detector as an example. For NS waveforms within the COMPAS-based DNS population, even though the SNR detected by the Cosmic Explorer decreases with increasing eccentricity, the orbital period of this population increases with eccentricity. Consequently, the parameter estimation for $\ln P_{\rm b}$, $\ln e$, $\cos \iota$, and $\theta_S$ improves with increasing eccentricity. There remains an open question on the number of detectable NSs emitting continuous GWs in eccentric orbits, which we leave for future studies.
We note that many current continuous wave searches use lower fitting factors due to computational limitations \cite{Mukherjee:2022tuc}. Also, torque effects limit coherent integration times to several days for accreting binary systems like Sco X-1 \cite{Zhang2021}. These practical considerations should be taken into account in future work.
 
\begin{acknowledgments}

We thank the anonymous referee for helpful comments and suggestions.
W.-F.F. is supported by the National Natural Science Foundation of China under Grant No. 12447109.
{T.L. is supported by the China Postdoctoral Science Foundation Grant No. 2024M760692.}
{Y.W. gratefully acknowledges support from the National Key Research and Development Program of China (No. 2023YFC2206702 and No. 2022YFC2205201), the National Natural Science Foundation of China under Grant No. 11973024, Major Science and Technology Program of Xinjiang Uygur Autonomous Region (No. 2022A03013-4), and Guangdong Major Project of Basic and Applied Basic Research (Grant No. 2019B030302001). }
{L.S. acknowledges support from the Beijing Natural Science Foundation (1242018), the National SKA Program of China (2020SKA0120300), the National Natural Science Foundation of China (11991053), the Max Planck Partner Group Program funded by the Max Planck Society, and the High-Performance Computing Platform of Peking University.}

\end{acknowledgments}

\bibliography{reference}

\begin{thebibliography}{61}%
\makeatletter
\providecommand \@ifxundefined [1]{%
 \@ifx{#1\undefined}
}%
\providecommand \@ifnum [1]{%
 \ifnum #1\expandafter \@firstoftwo
 \else \expandafter \@secondoftwo
 \fi
}%
\providecommand \@ifx [1]{%
 \ifx #1\expandafter \@firstoftwo
 \else \expandafter \@secondoftwo
 \fi
}%
\providecommand \natexlab [1]{#1}%
\providecommand \enquote  [1]{``#1''}%
\providecommand \bibnamefont  [1]{#1}%
\providecommand \bibfnamefont [1]{#1}%
\providecommand \citenamefont [1]{#1}%
\providecommand \href@noop [0]{\@secondoftwo}%
\providecommand \href [0]{\begingroup \@sanitize@url \@href}%
\providecommand \@href[1]{\@@startlink{#1}\@@href}%
\providecommand \@@href[1]{\endgroup#1\@@endlink}%
\providecommand \@sanitize@url [0]{\catcode `\\12\catcode `\$12\catcode
  `\&12\catcode `\#12\catcode `\^12\catcode `\_12\catcode `\%12\relax}%
\providecommand \@@startlink[1]{}%
\providecommand \@@endlink[0]{}%
\providecommand \url  [0]{\begingroup\@sanitize@url \@url }%
\providecommand \@url [1]{\endgroup\@href {#1}{\urlprefix }}%
\providecommand \urlprefix  [0]{URL }%
\providecommand \Eprint [0]{\href }%
\providecommand \doibase [0]{http://dx.doi.org/}%
\providecommand \selectlanguage [0]{\@gobble}%
\providecommand \bibinfo  [0]{\@secondoftwo}%
\providecommand \bibfield  [0]{\@secondoftwo}%
\providecommand \translation [1]{[#1]}%
\providecommand \BibitemOpen [0]{}%
\providecommand \bibitemStop [0]{}%
\providecommand \bibitemNoStop [0]{.\EOS\space}%
\providecommand \EOS [0]{\spacefactor3000\relax}%
\providecommand \BibitemShut  [1]{\csname bibitem#1\endcsname}%
\let\auto@bib@innerbib\@empty
\bibitem [{\citenamefont {{Sieniawska}}\ and\ \citenamefont
  {{Bejger}}(2019)}]{Sieniawska2019}%
  \BibitemOpen
  \bibfield  {author} {\bibinfo {author} {\bibfnamefont {M.}~\bibnamefont
  {{Sieniawska}}}\ and\ \bibinfo {author} {\bibfnamefont {M.}~\bibnamefont
  {{Bejger}}},\ }\href {\doibase 10.3390/universe5110217} {\bibfield  {journal}
  {\bibinfo  {journal} {Universe}\ }\textbf {\bibinfo {volume} {5}},\ \bibinfo
  {pages} {217} (\bibinfo {year} {2019})},\ \Eprint
  {http://arxiv.org/abs/1909.12600} {arXiv:1909.12600 [astro-ph.HE]}
  \BibitemShut {NoStop}%
\bibitem [{\citenamefont {{Riles}}(2017)}]{Riles2017}%
  \BibitemOpen
  \bibfield  {author} {\bibinfo {author} {\bibfnamefont {K.}~\bibnamefont
  {{Riles}}},\ }\href {\doibase 10.1142/S021773231730035X} {\bibfield
  {journal} {\bibinfo  {journal} {Mod. Phys. Lett. A}\ }\textbf {\bibinfo
  {volume} {32}},\ \bibinfo {eid} {1730035-685} (\bibinfo {year} {2017})},\
  \Eprint {http://arxiv.org/abs/1712.05897} {arXiv:1712.05897 [gr-qc]}
  \BibitemShut {NoStop}%
\bibitem [{\citenamefont {{Lasky}}(2015)}]{Lasky2015}%
  \BibitemOpen
  \bibfield  {author} {\bibinfo {author} {\bibfnamefont {P.~D.}\ \bibnamefont
  {{Lasky}}},\ }\href {\doibase 10.1017/pasa.2015.35} {\bibfield  {journal}
  {\bibinfo  {journal} {Publ. Astron. Soc. Aust.}\ }\textbf {\bibinfo {volume}
  {32}},\ \bibinfo {eid} {e034} (\bibinfo {year} {2015})},\ \Eprint
  {http://arxiv.org/abs/1508.06643} {arXiv:1508.06643 [astro-ph.HE]}
  \BibitemShut {NoStop}%
\bibitem [{\citenamefont {Riles}(2023)}]{2022arXiv220606447R}%
  \BibitemOpen
  \bibfield  {author} {\bibinfo {author} {\bibfnamefont {K.}~\bibnamefont
  {Riles}},\ }\href {\doibase 10.1007/s41114-023-00044-3} {\bibfield  {journal}
  {\bibinfo  {journal} {Living Rev. Rel.}\ }\textbf {\bibinfo {volume} {26}},\
  \bibinfo {pages} {3} (\bibinfo {year} {2023})},\ \Eprint
  {http://arxiv.org/abs/2206.06447} {arXiv:2206.06447 [astro-ph.HE]}
  \BibitemShut {NoStop}%
\bibitem [{\citenamefont {Wette}(2023)}]{Wette2023}%
  \BibitemOpen
  \bibfield  {author} {\bibinfo {author} {\bibfnamefont {K.}~\bibnamefont
  {Wette}},\ }\href {\doibase 10.1016/j.astropartphys.2023.102880} {\bibfield
  {journal} {\bibinfo  {journal} {Astropart. Phys.}\ }\textbf {\bibinfo
  {volume} {153}},\ \bibinfo {pages} {102880} (\bibinfo {year} {2023})},\
  \Eprint {http://arxiv.org/abs/2305.07106} {arXiv:2305.07106 [gr-qc]}
  \BibitemShut {NoStop}%
\bibitem [{\citenamefont {{Pitkin}}(2011)}]{Pitkin2011}%
  \BibitemOpen
  \bibfield  {author} {\bibinfo {author} {\bibfnamefont {M.}~\bibnamefont
  {{Pitkin}}},\ }\href {\doibase 10.1111/j.1365-2966.2011.18818.x} {\bibfield
  {journal} {\bibinfo  {journal} {Mon. Not. Roy. Astron. Soc.}\ }\textbf
  {\bibinfo {volume} {415}},\ \bibinfo {pages} {1849} (\bibinfo {year}
  {2011})},\ \Eprint {http://arxiv.org/abs/1103.5867} {arXiv:1103.5867
  [astro-ph.HE]} \BibitemShut {NoStop}%
\bibitem [{\citenamefont {{Soldateschi}}\ and\ \citenamefont
  {{Bucciantini}}(2021)}]{Soldateschi2021}%
  \BibitemOpen
  \bibfield  {author} {\bibinfo {author} {\bibfnamefont {J.}~\bibnamefont
  {{Soldateschi}}}\ and\ \bibinfo {author} {\bibfnamefont {N.}~\bibnamefont
  {{Bucciantini}}},\ }\href {\doibase 10.3390/galaxies9040101} {\bibfield
  {journal} {\bibinfo  {journal} {Galaxies}\ }\textbf {\bibinfo {volume} {9}},\
  \bibinfo {pages} {101} (\bibinfo {year} {2021})},\ \Eprint
  {http://arxiv.org/abs/2110.06039} {arXiv:2110.06039 [astro-ph.HE]}
  \BibitemShut {NoStop}%
\bibitem [{\citenamefont {Lu}\ \emph {et~al.}(2023)\citenamefont {Lu},
  \citenamefont {Wette}, \citenamefont {Scott},\ and\ \citenamefont
  {Melatos}}]{Lu2022}%
  \BibitemOpen
  \bibfield  {author} {\bibinfo {author} {\bibfnamefont {N.}~\bibnamefont
  {Lu}}, \bibinfo {author} {\bibfnamefont {K.}~\bibnamefont {Wette}}, \bibinfo
  {author} {\bibfnamefont {S.~M.}\ \bibnamefont {Scott}}, \ and\ \bibinfo
  {author} {\bibfnamefont {A.}~\bibnamefont {Melatos}},\ }\href {\doibase
  10.1093/mnras/stad390} {\bibfield  {journal} {\bibinfo  {journal} {Mon. Not.
  R. Astron. Soc.}\ }\textbf {\bibinfo {volume} {521}},\ \bibinfo {pages}
  {2103} (\bibinfo {year} {2023})}\BibitemShut {NoStop}%
\bibitem [{\citenamefont {Yim}\ \emph {et~al.}(2023)\citenamefont {Yim},
  \citenamefont {Gao}, \citenamefont {Kang}, \citenamefont {Shao},\ and\
  \citenamefont {Xu}}]{Yim:2023nda}%
  \BibitemOpen
  \bibfield  {author} {\bibinfo {author} {\bibfnamefont {G.}~\bibnamefont
  {Yim}}, \bibinfo {author} {\bibfnamefont {Y.}~\bibnamefont {Gao}}, \bibinfo
  {author} {\bibfnamefont {Y.}~\bibnamefont {Kang}}, \bibinfo {author}
  {\bibfnamefont {L.}~\bibnamefont {Shao}}, \ and\ \bibinfo {author}
  {\bibfnamefont {R.}~\bibnamefont {Xu}},\ }\href {\doibase
  10.1093/mnras/stad3337} {\bibfield  {journal} {\bibinfo  {journal} {Mon. Not.
  Roy. Astron. Soc.}\ }\textbf {\bibinfo {volume} {527}},\ \bibinfo {pages}
  {2379} (\bibinfo {year} {2023})},\ \Eprint {http://arxiv.org/abs/2308.01588}
  {arXiv:2308.01588 [astro-ph.HE]} \BibitemShut {NoStop}%
\bibitem [{\citenamefont {Feng}\ \emph {et~al.}(2024)\citenamefont {Feng},
  \citenamefont {Chen}, \citenamefont {Liu}, \citenamefont {Wang},\ and\
  \citenamefont {Mohanty}}]{Feng2024}%
  \BibitemOpen
  \bibfield  {author} {\bibinfo {author} {\bibfnamefont {W.-F.}\ \bibnamefont
  {Feng}}, \bibinfo {author} {\bibfnamefont {J.-W.}\ \bibnamefont {Chen}},
  \bibinfo {author} {\bibfnamefont {T.}~\bibnamefont {Liu}}, \bibinfo {author}
  {\bibfnamefont {Y.}~\bibnamefont {Wang}}, \ and\ \bibinfo {author}
  {\bibfnamefont {S.~D.}\ \bibnamefont {Mohanty}},\ }\href {\doibase
  10.1103/PhysRevD.109.043033} {\bibfield  {journal} {\bibinfo  {journal}
  {Phys. Rev. D}\ }\textbf {\bibinfo {volume} {109}},\ \bibinfo {pages}
  {043033} (\bibinfo {year} {2024})}\BibitemShut {NoStop}%
\bibitem [{\citenamefont {Aasi}\ \emph {et~al.}(2015)\citenamefont {Aasi} \emph
  {et~al.}}]{advancedLIGO2015}%
  \BibitemOpen
  \bibfield  {author} {\bibinfo {author} {\bibfnamefont {J.}~\bibnamefont
  {Aasi}} \emph {et~al.} (\bibinfo {collaboration} {LIGO Scientific}),\ }\href
  {\doibase 10.1088/0264-9381/32/7/074001} {\bibfield  {journal} {\bibinfo
  {journal} {Class. Quant. Grav.}\ }\textbf {\bibinfo {volume} {32}},\ \bibinfo
  {pages} {074001} (\bibinfo {year} {2015})},\ \Eprint
  {http://arxiv.org/abs/1411.4547} {arXiv:1411.4547 [gr-qc]} \BibitemShut
  {NoStop}%
\bibitem [{\citenamefont {Acernese}\ \emph {et~al.}(2015)\citenamefont
  {Acernese} \emph {et~al.}}]{AdvancedVirgo2015}%
  \BibitemOpen
  \bibfield  {author} {\bibinfo {author} {\bibfnamefont {F.}~\bibnamefont
  {Acernese}} \emph {et~al.} (\bibinfo {collaboration} {VIRGO}),\ }\href
  {\doibase 10.1088/0264-9381/32/2/024001} {\bibfield  {journal} {\bibinfo
  {journal} {Class. Quant. Grav.}\ }\textbf {\bibinfo {volume} {32}},\ \bibinfo
  {pages} {024001} (\bibinfo {year} {2015})},\ \Eprint
  {http://arxiv.org/abs/1408.3978} {arXiv:1408.3978 [gr-qc]} \BibitemShut
  {NoStop}%
\bibitem [{\citenamefont {Akutsu}\ \emph {et~al.}(2020)\citenamefont {Akutsu}
  \emph {et~al.}}]{KAGRA}%
  \BibitemOpen
  \bibfield  {author} {\bibinfo {author} {\bibfnamefont {T.}~\bibnamefont
  {Akutsu}} \emph {et~al.},\ }\href {\doibase 10.1093/ptep/ptaa125} {\bibfield
  {journal} {\bibinfo  {journal} {Prog. Theor. Exp. Phys.}\ }\textbf {\bibinfo
  {volume} {2021}},\ \bibinfo {pages} {05A101} (\bibinfo {year}
  {2020})}\BibitemShut {NoStop}%
\bibitem [{\citenamefont {Abbott}\ \emph {et~al.}(2022)\citenamefont {Abbott}
  \emph {et~al.}}]{LIGO2022isolatedO3}%
  \BibitemOpen
  \bibfield  {author} {\bibinfo {author} {\bibfnamefont {R.}~\bibnamefont
  {Abbott}} \emph {et~al.} (\bibinfo {collaboration} {KAGRA, LIGO Scientific,
  VIRGO}),\ }\href {\doibase 10.1103/PhysRevD.106.102008} {\bibfield  {journal}
  {\bibinfo  {journal} {Phys. Rev. D}\ }\textbf {\bibinfo {volume} {106}},\
  \bibinfo {pages} {102008} (\bibinfo {year} {2022})},\ \Eprint
  {http://arxiv.org/abs/2201.00697} {arXiv:2201.00697 [gr-qc]} \BibitemShut
  {NoStop}%
\bibitem [{\citenamefont {Abbott}\ \emph
  {et~al.}(2021{\natexlab{a}})\citenamefont {Abbott}, \citenamefont {Abbott},\
  and\ \citenamefont {et~al.}}]{LIGOisolatedNS2021}%
  \BibitemOpen
  \bibfield  {author} {\bibinfo {author} {\bibfnamefont {R.}~\bibnamefont
  {Abbott}}, \bibinfo {author} {\bibfnamefont {T.~D.}\ \bibnamefont {Abbott}},
  \ and\ \bibinfo {author} {\bibnamefont {et~al.}} (\bibinfo {collaboration}
  {LIGO Scientific Collaboration, Virgo Collaboration, and KAGRA
  Collaboration}),\ }\href {\doibase 10.1103/PhysRevD.104.082004} {\bibfield
  {journal} {\bibinfo  {journal} {Phys. Rev. D}\ }\textbf {\bibinfo {volume}
  {104}},\ \bibinfo {pages} {082004} (\bibinfo {year}
  {2021}{\natexlab{a}})}\BibitemShut {NoStop}%
\bibitem [{\citenamefont {{Abbott}}\ \emph
  {et~al.}(2022{\natexlab{a}})\citenamefont {{Abbott}}, \citenamefont
  {{Abbott}},\ and\ \citenamefont {et~al.}}]{Abbott2022Narrowband}%
  \BibitemOpen
  \bibfield  {author} {\bibinfo {author} {\bibfnamefont {R.}~\bibnamefont
  {{Abbott}}}, \bibinfo {author} {\bibfnamefont {T.~D.}\ \bibnamefont
  {{Abbott}}}, \ and\ \bibinfo {author} {\bibnamefont {et~al.}},\ }\href
  {\doibase 10.3847/1538-4357/ac6ad0} {\bibfield  {journal} {\bibinfo
  {journal} {\apj}\ }\textbf {\bibinfo {volume} {932}},\ \bibinfo {eid} {133}
  (\bibinfo {year} {2022}{\natexlab{a}})},\ \Eprint
  {http://arxiv.org/abs/2112.10990} {arXiv:2112.10990 [gr-qc]} \BibitemShut
  {NoStop}%
\bibitem [{\citenamefont {{Abbott}}\ \emph
  {et~al.}(2022{\natexlab{b}})\citenamefont {{Abbott}}, \citenamefont
  {{Abbott}},\ and\ \citenamefont {et~al.}}]{Abbott2022CasA}%
  \BibitemOpen
  \bibfield  {author} {\bibinfo {author} {\bibfnamefont {R.}~\bibnamefont
  {{Abbott}}}, \bibinfo {author} {\bibfnamefont {T.~D.}\ \bibnamefont
  {{Abbott}}}, \ and\ \bibinfo {author} {\bibnamefont {et~al.}},\ }\href
  {\doibase 10.1103/PhysRevD.105.082005} {\bibfield  {journal} {\bibinfo
  {journal} {\prd}\ }\textbf {\bibinfo {volume} {105}},\ \bibinfo {eid}
  {082005} (\bibinfo {year} {2022}{\natexlab{b}})},\ \Eprint
  {http://arxiv.org/abs/2111.15116} {arXiv:2111.15116 [gr-qc]} \BibitemShut
  {NoStop}%
\bibitem [{\citenamefont {Abbott}\ \emph
  {et~al.}(2021{\natexlab{b}})\citenamefont {Abbott}, \citenamefont {Abbott},\
  and\ \citenamefont {et~al.}}]{LIGONSinBinary2021}%
  \BibitemOpen
  \bibfield  {author} {\bibinfo {author} {\bibfnamefont {R.}~\bibnamefont
  {Abbott}}, \bibinfo {author} {\bibfnamefont {T.~D.}\ \bibnamefont {Abbott}},
  \ and\ \bibinfo {author} {\bibnamefont {et~al.}} (\bibinfo {collaboration}
  {The LIGO Scientific Collaboration and the Virgo Collaboration}),\ }\href
  {\doibase 10.1103/PhysRevD.103.064017} {\bibfield  {journal} {\bibinfo
  {journal} {\prd}\ }\textbf {\bibinfo {volume} {103}},\ \bibinfo {pages}
  {064017} (\bibinfo {year} {2021}{\natexlab{b}})}\BibitemShut {NoStop}%
\bibitem [{\citenamefont {{Covas}}\ and\ \citenamefont
  {{Sintes}}(2020)}]{Covas2020}%
  \BibitemOpen
  \bibfield  {author} {\bibinfo {author} {\bibfnamefont {P.~B.}\ \bibnamefont
  {{Covas}}}\ and\ \bibinfo {author} {\bibfnamefont {A.~M.}\ \bibnamefont
  {{Sintes}}},\ }\href {\doibase 10.1103/PhysRevLett.124.191102} {\bibfield
  {journal} {\bibinfo  {journal} {\prl}\ }\textbf {\bibinfo {volume} {124}},\
  \bibinfo {eid} {191102} (\bibinfo {year} {2020})},\ \Eprint
  {http://arxiv.org/abs/2001.08411} {arXiv:2001.08411 [gr-qc]} \BibitemShut
  {NoStop}%
\bibitem [{\citenamefont {{Covas}}\ and\ \citenamefont
  {{Sintes}}(2019)}]{Covas2019}%
  \BibitemOpen
  \bibfield  {author} {\bibinfo {author} {\bibfnamefont {P.~B.}\ \bibnamefont
  {{Covas}}}\ and\ \bibinfo {author} {\bibfnamefont {A.~M.}\ \bibnamefont
  {{Sintes}}},\ }\href {\doibase 10.1103/PhysRevD.99.124019} {\bibfield
  {journal} {\bibinfo  {journal} {\prd}\ }\textbf {\bibinfo {volume} {99}},\
  \bibinfo {eid} {124019} (\bibinfo {year} {2019})},\ \Eprint
  {http://arxiv.org/abs/1904.04873} {arXiv:1904.04873 [astro-ph.IM]}
  \BibitemShut {NoStop}%
\bibitem [{\citenamefont {Zhang}\ \emph {et~al.}(2021)\citenamefont {Zhang},
  \citenamefont {Papa}, \citenamefont {Krishnan},\ and\ \citenamefont
  {Watts}}]{Zhang2021}%
  \BibitemOpen
  \bibfield  {author} {\bibinfo {author} {\bibfnamefont {Y.}~\bibnamefont
  {Zhang}}, \bibinfo {author} {\bibfnamefont {M.~A.}\ \bibnamefont {Papa}},
  \bibinfo {author} {\bibfnamefont {B.}~\bibnamefont {Krishnan}}, \ and\
  \bibinfo {author} {\bibfnamefont {A.~L.}\ \bibnamefont {Watts}},\ }\href
  {\doibase 10.3847/2041-8213/abd256} {\bibfield  {journal} {\bibinfo
  {journal} {Astrophys. J. Lett.}\ }\textbf {\bibinfo {volume} {906}},\
  \bibinfo {pages} {L14} (\bibinfo {year} {2021})},\ \Eprint
  {http://arxiv.org/abs/2011.04414} {arXiv:2011.04414 [astro-ph.HE]}
  \BibitemShut {NoStop}%
\bibitem [{\citenamefont {{Leaci}}\ and\ \citenamefont
  {{Prix}}(2015)}]{Leaci2015}%
  \BibitemOpen
  \bibfield  {author} {\bibinfo {author} {\bibfnamefont {P.}~\bibnamefont
  {{Leaci}}}\ and\ \bibinfo {author} {\bibfnamefont {R.}~\bibnamefont
  {{Prix}}},\ }\href {\doibase 10.1103/PhysRevD.91.102003} {\bibfield
  {journal} {\bibinfo  {journal} {\prd}\ }\textbf {\bibinfo {volume} {91}},\
  \bibinfo {eid} {102003} (\bibinfo {year} {2015})},\ \Eprint
  {http://arxiv.org/abs/1502.00914} {arXiv:1502.00914 [gr-qc]} \BibitemShut
  {NoStop}%
\bibitem [{\citenamefont {Pagliaro}\ \emph {et~al.}(2023)\citenamefont
  {Pagliaro}, \citenamefont {Papa}, \citenamefont {Ming}, \citenamefont {Lian},
  \citenamefont {Tsuna}, \citenamefont {Maraston},\ and\ \citenamefont
  {Thomas}}]{Pagliaro:2023bvi}%
  \BibitemOpen
  \bibfield  {author} {\bibinfo {author} {\bibfnamefont {G.}~\bibnamefont
  {Pagliaro}}, \bibinfo {author} {\bibfnamefont {M.~A.}\ \bibnamefont {Papa}},
  \bibinfo {author} {\bibfnamefont {J.}~\bibnamefont {Ming}}, \bibinfo {author}
  {\bibfnamefont {J.}~\bibnamefont {Lian}}, \bibinfo {author} {\bibfnamefont
  {D.}~\bibnamefont {Tsuna}}, \bibinfo {author} {\bibfnamefont
  {C.}~\bibnamefont {Maraston}}, \ and\ \bibinfo {author} {\bibfnamefont
  {D.}~\bibnamefont {Thomas}},\ }\href {\doibase 10.3847/1538-4357/acd76f}
  {\bibfield  {journal} {\bibinfo  {journal} {Astrophys. J.}\ }\textbf
  {\bibinfo {volume} {952}},\ \bibinfo {pages} {123} (\bibinfo {year}
  {2023})},\ \Eprint {http://arxiv.org/abs/2303.04714} {arXiv:2303.04714
  [gr-qc]} \BibitemShut {NoStop}%
\bibitem [{\citenamefont {{Covas}}\ \emph {et~al.}(2024)\citenamefont
  {{Covas}}, \citenamefont {{Papa}},\ and\ \citenamefont
  {{Prix}}}]{Covas:2024nzs}%
  \BibitemOpen
  \bibfield  {author} {\bibinfo {author} {\bibfnamefont {P.~B.}\ \bibnamefont
  {{Covas}}}, \bibinfo {author} {\bibfnamefont {M.~A.}\ \bibnamefont {{Papa}}},
  \ and\ \bibinfo {author} {\bibfnamefont {R.}~\bibnamefont {{Prix}}},\ }\href
  {\doibase 10.48550/arXiv.2409.16196} {\bibfield  {journal} {\bibinfo
  {journal} {arXiv e-prints}\ ,\ \bibinfo {eid} {arXiv:2409.16196}} (\bibinfo
  {year} {2024})},\ \Eprint {http://arxiv.org/abs/2409.16196} {arXiv:2409.16196
  [gr-qc]} \BibitemShut {NoStop}%
\bibitem [{\citenamefont {{Ming}}\ \emph {et~al.}(2024)\citenamefont {{Ming}},
  \citenamefont {{Alessandra Papa}}, \citenamefont {{Eggenstein}},
  \citenamefont {{Beheshtipour}}, \citenamefont {{Machenschalk}}, \citenamefont
  {{Prix}}, \citenamefont {{Allen}},\ and\ \citenamefont
  {{Bensch}}}]{Ming2024}%
  \BibitemOpen
  \bibfield  {author} {\bibinfo {author} {\bibfnamefont {J.}~\bibnamefont
  {{Ming}}}, \bibinfo {author} {\bibfnamefont {M.}~\bibnamefont {{Alessandra
  Papa}}}, \bibinfo {author} {\bibfnamefont {H.-B.}\ \bibnamefont
  {{Eggenstein}}}, \bibinfo {author} {\bibfnamefont {B.}~\bibnamefont
  {{Beheshtipour}}}, \bibinfo {author} {\bibfnamefont {B.}~\bibnamefont
  {{Machenschalk}}}, \bibinfo {author} {\bibfnamefont {R.}~\bibnamefont
  {{Prix}}}, \bibinfo {author} {\bibfnamefont {B.}~\bibnamefont {{Allen}}}, \
  and\ \bibinfo {author} {\bibfnamefont {M.}~\bibnamefont {{Bensch}}},\ }\href
  {\doibase 10.48550/arXiv.2408.14573} {\bibfield  {journal} {\bibinfo
  {journal} {arXiv e-prints}\ ,\ \bibinfo {eid} {arXiv:2408.14573}} (\bibinfo
  {year} {2024})},\ \Eprint {http://arxiv.org/abs/2408.14573} {arXiv:2408.14573
  [gr-qc]} \BibitemShut {NoStop}%
\bibitem [{\citenamefont {{Owen}}\ \emph {et~al.}(2024)\citenamefont {{Owen}},
  \citenamefont {{Lindblom}}, \citenamefont {{Soares Pinheiro}},\ and\
  \citenamefont {{Rajbhandari}}}]{Owen2024}%
  \BibitemOpen
  \bibfield  {author} {\bibinfo {author} {\bibfnamefont {B.~J.}\ \bibnamefont
  {{Owen}}}, \bibinfo {author} {\bibfnamefont {L.}~\bibnamefont {{Lindblom}}},
  \bibinfo {author} {\bibfnamefont {L.}~\bibnamefont {{Soares Pinheiro}}}, \
  and\ \bibinfo {author} {\bibfnamefont {B.}~\bibnamefont {{Rajbhandari}}},\
  }\href {\doibase 10.3847/2041-8213/ad2263} {\bibfield  {journal} {\bibinfo
  {journal} {Astrophys. J. Lett.}\ }\textbf {\bibinfo {volume} {962}},\
  \bibinfo {eid} {L23} (\bibinfo {year} {2024})},\ \Eprint
  {http://arxiv.org/abs/2310.19964} {arXiv:2310.19964 [gr-qc]} \BibitemShut
  {NoStop}%
\bibitem [{\citenamefont {{Liu}}\ and\ \citenamefont
  {{Zou}}(2022)}]{LiuandZou2022}%
  \BibitemOpen
  \bibfield  {author} {\bibinfo {author} {\bibfnamefont {Y.}~\bibnamefont
  {{Liu}}}\ and\ \bibinfo {author} {\bibfnamefont {Y.-C.}\ \bibnamefont
  {{Zou}}},\ }\href {\doibase 10.1103/PhysRevD.106.123024} {\bibfield
  {journal} {\bibinfo  {journal} {\prd}\ }\textbf {\bibinfo {volume} {106}},\
  \bibinfo {eid} {123024} (\bibinfo {year} {2022})},\ \Eprint
  {http://arxiv.org/abs/2211.02855} {arXiv:2211.02855 [astro-ph.HE]}
  \BibitemShut {NoStop}%
\bibitem [{\citenamefont {{Rajbhandari}}\ \emph {et~al.}(2021)\citenamefont
  {{Rajbhandari}}, \citenamefont {{Owen}}, \citenamefont {{Caride}},\ and\
  \citenamefont {{Inta}}}]{Rajbhandari2021}%
  \BibitemOpen
  \bibfield  {author} {\bibinfo {author} {\bibfnamefont {B.}~\bibnamefont
  {{Rajbhandari}}}, \bibinfo {author} {\bibfnamefont {B.~J.}\ \bibnamefont
  {{Owen}}}, \bibinfo {author} {\bibfnamefont {S.}~\bibnamefont {{Caride}}}, \
  and\ \bibinfo {author} {\bibfnamefont {R.}~\bibnamefont {{Inta}}},\ }\href
  {\doibase 10.1103/PhysRevD.104.122008} {\bibfield  {journal} {\bibinfo
  {journal} {\prd}\ }\textbf {\bibinfo {volume} {104}},\ \bibinfo {eid}
  {122008} (\bibinfo {year} {2021})},\ \Eprint
  {http://arxiv.org/abs/2101.00714} {arXiv:2101.00714 [gr-qc]} \BibitemShut
  {NoStop}%
\bibitem [{\citenamefont {{Dergachev}}\ \emph {et~al.}(2019)\citenamefont
  {{Dergachev}}, \citenamefont {{Papa}}, \citenamefont {{Steltner}},\ and\
  \citenamefont {{Eggenstein}}}]{Dergachev2019}%
  \BibitemOpen
  \bibfield  {author} {\bibinfo {author} {\bibfnamefont {V.}~\bibnamefont
  {{Dergachev}}}, \bibinfo {author} {\bibfnamefont {M.~A.}\ \bibnamefont
  {{Papa}}}, \bibinfo {author} {\bibfnamefont {B.}~\bibnamefont {{Steltner}}},
  \ and\ \bibinfo {author} {\bibfnamefont {H.-B.}\ \bibnamefont
  {{Eggenstein}}},\ }\href {\doibase 10.1103/PhysRevD.99.084048} {\bibfield
  {journal} {\bibinfo  {journal} {\prd}\ }\textbf {\bibinfo {volume} {99}},\
  \bibinfo {eid} {084048} (\bibinfo {year} {2019})},\ \Eprint
  {http://arxiv.org/abs/1903.02389} {arXiv:1903.02389 [gr-qc]} \BibitemShut
  {NoStop}%
\bibitem [{\citenamefont {{Zhu}}\ \emph {et~al.}(2016)\citenamefont {{Zhu}},
  \citenamefont {{Papa}}, \citenamefont {{Eggenstein}}, \citenamefont {{Prix}},
  \citenamefont {{Wette}}, \citenamefont {{Allen}}, \citenamefont {{Bock}},
  \citenamefont {{Keitel}}, \citenamefont {{Krishnan}}, \citenamefont
  {{Machenschalk}}, \citenamefont {{Shaltev}},\ and\ \citenamefont
  {{Siemens}}}]{Zhu2016}%
  \BibitemOpen
  \bibfield  {author} {\bibinfo {author} {\bibfnamefont {S.~J.}\ \bibnamefont
  {{Zhu}}}, \bibinfo {author} {\bibfnamefont {M.~A.}\ \bibnamefont {{Papa}}},
  \bibinfo {author} {\bibfnamefont {H.-B.}\ \bibnamefont {{Eggenstein}}},
  \bibinfo {author} {\bibfnamefont {R.}~\bibnamefont {{Prix}}}, \bibinfo
  {author} {\bibfnamefont {K.}~\bibnamefont {{Wette}}}, \bibinfo {author}
  {\bibfnamefont {B.}~\bibnamefont {{Allen}}}, \bibinfo {author} {\bibfnamefont
  {O.}~\bibnamefont {{Bock}}}, \bibinfo {author} {\bibfnamefont
  {D.}~\bibnamefont {{Keitel}}}, \bibinfo {author} {\bibfnamefont
  {B.}~\bibnamefont {{Krishnan}}}, \bibinfo {author} {\bibfnamefont
  {B.}~\bibnamefont {{Machenschalk}}}, \bibinfo {author} {\bibfnamefont
  {M.}~\bibnamefont {{Shaltev}}}, \ and\ \bibinfo {author} {\bibfnamefont
  {X.}~\bibnamefont {{Siemens}}},\ }\href {\doibase 10.1103/PhysRevD.94.082008}
  {\bibfield  {journal} {\bibinfo  {journal} {\prd}\ }\textbf {\bibinfo
  {volume} {94}},\ \bibinfo {eid} {082008} (\bibinfo {year} {2016})},\ \Eprint
  {http://arxiv.org/abs/1608.07589} {arXiv:1608.07589 [gr-qc]} \BibitemShut
  {NoStop}%
\bibitem [{\citenamefont {Amaro-Seoane}\ \emph {et~al.}(2017)\citenamefont
  {Amaro-Seoane} \emph {et~al.}}]{LISA2017}%
  \BibitemOpen
  \bibfield  {author} {\bibinfo {author} {\bibfnamefont {P.}~\bibnamefont
  {Amaro-Seoane}} \emph {et~al.} (\bibinfo {collaboration} {LISA}),\
  }\href@noop {} {\  (\bibinfo {year} {2017})},\ \Eprint
  {http://arxiv.org/abs/1702.00786} {arXiv:1702.00786 [astro-ph.IM]}
  \BibitemShut {NoStop}%
\bibitem [{\citenamefont {Luo}\ \emph {et~al.}(2016)\citenamefont {Luo} \emph
  {et~al.}}]{TianQin2016}%
  \BibitemOpen
  \bibfield  {author} {\bibinfo {author} {\bibfnamefont {J.}~\bibnamefont
  {Luo}} \emph {et~al.} (\bibinfo {collaboration} {TianQin}),\ }\href {\doibase
  10.1088/0264-9381/33/3/035010} {\bibfield  {journal} {\bibinfo  {journal}
  {Class. Quant. Grav.}\ }\textbf {\bibinfo {volume} {33}},\ \bibinfo {pages}
  {035010} (\bibinfo {year} {2016})},\ \Eprint
  {http://arxiv.org/abs/1512.02076} {arXiv:1512.02076 [astro-ph.IM]}
  \BibitemShut {NoStop}%
\bibitem [{\citenamefont {Hu}\ and\ \citenamefont {Wu}(2017)}]{Taiji2017}%
  \BibitemOpen
  \bibfield  {author} {\bibinfo {author} {\bibfnamefont {W.-R.}\ \bibnamefont
  {Hu}}\ and\ \bibinfo {author} {\bibfnamefont {Y.-L.}\ \bibnamefont {Wu}},\
  }\href {\doibase 10.1093/nsr/nwx116} {\bibfield  {journal} {\bibinfo
  {journal} {Natl. Sci. Rev.}\ }\textbf {\bibinfo {volume} {4}},\ \bibinfo
  {pages} {685} (\bibinfo {year} {2017})}\BibitemShut {NoStop}%
\bibitem [{\citenamefont {Andrews}\ \emph {et~al.}(2020)\citenamefont
  {Andrews}, \citenamefont {Breivik}, \citenamefont {Pankow}, \citenamefont
  {D'Orazio},\ and\ \citenamefont {Safarzadeh}}]{Andrews2020}%
  \BibitemOpen
  \bibfield  {author} {\bibinfo {author} {\bibfnamefont {J.~J.}\ \bibnamefont
  {Andrews}}, \bibinfo {author} {\bibfnamefont {K.}~\bibnamefont {Breivik}},
  \bibinfo {author} {\bibfnamefont {C.}~\bibnamefont {Pankow}}, \bibinfo
  {author} {\bibfnamefont {D.~J.}\ \bibnamefont {D'Orazio}}, \ and\ \bibinfo
  {author} {\bibfnamefont {M.}~\bibnamefont {Safarzadeh}},\ }\href {\doibase
  10.3847/2041-8213/ab5b9a} {\bibfield  {journal} {\bibinfo  {journal}
  {Astrophys. J. Lett.}\ }\textbf {\bibinfo {volume} {892}},\ \bibinfo {pages}
  {L9} (\bibinfo {year} {2020})},\ \Eprint {http://arxiv.org/abs/1910.13436}
  {arXiv:1910.13436 [astro-ph.HE]} \BibitemShut {NoStop}%
\bibitem [{\citenamefont {Miao}\ \emph {et~al.}(2021)\citenamefont {Miao},
  \citenamefont {Xu}, \citenamefont {Shao}, \citenamefont {Liu},\ and\
  \citenamefont {Ma}}]{Miao:2021awa}%
  \BibitemOpen
  \bibfield  {author} {\bibinfo {author} {\bibfnamefont {X.}~\bibnamefont
  {Miao}}, \bibinfo {author} {\bibfnamefont {H.}~\bibnamefont {Xu}}, \bibinfo
  {author} {\bibfnamefont {L.}~\bibnamefont {Shao}}, \bibinfo {author}
  {\bibfnamefont {C.}~\bibnamefont {Liu}}, \ and\ \bibinfo {author}
  {\bibfnamefont {B.-Q.}\ \bibnamefont {Ma}},\ }\href {\doibase
  10.3847/1538-4357/ac1d48} {\bibfield  {journal} {\bibinfo  {journal}
  {Astrophys. J.}\ }\textbf {\bibinfo {volume} {921}},\ \bibinfo {pages} {114}
  (\bibinfo {year} {2021})},\ \Eprint {http://arxiv.org/abs/2107.05812}
  {arXiv:2107.05812 [astro-ph.HE]} \BibitemShut {NoStop}%
\bibitem [{\citenamefont {{Wagg}}\ \emph {et~al.}(2022)\citenamefont {{Wagg}},
  \citenamefont {{Broekgaarden}}, \citenamefont {{de Mink}}, \citenamefont
  {{Frankel}}, \citenamefont {{van Son}},\ and\ \citenamefont
  {{Justham}}}]{Wagg2022}%
  \BibitemOpen
  \bibfield  {author} {\bibinfo {author} {\bibfnamefont {T.}~\bibnamefont
  {{Wagg}}}, \bibinfo {author} {\bibfnamefont {F.~S.}\ \bibnamefont
  {{Broekgaarden}}}, \bibinfo {author} {\bibfnamefont {S.~E.}\ \bibnamefont
  {{de Mink}}}, \bibinfo {author} {\bibfnamefont {N.}~\bibnamefont
  {{Frankel}}}, \bibinfo {author} {\bibfnamefont {L.~A.~C.}\ \bibnamefont {{van
  Son}}}, \ and\ \bibinfo {author} {\bibfnamefont {S.}~\bibnamefont
  {{Justham}}},\ }\href {\doibase 10.3847/1538-4357/ac8675} {\bibfield
  {journal} {\bibinfo  {journal} {\apj}\ }\textbf {\bibinfo {volume} {937}},\
  \bibinfo {eid} {118} (\bibinfo {year} {2022})},\ \Eprint
  {http://arxiv.org/abs/2111.13704} {arXiv:2111.13704 [astro-ph.HE]}
  \BibitemShut {NoStop}%
\bibitem [{\citenamefont {Feng}\ \emph
  {et~al.}(2023{\natexlab{a}})\citenamefont {Feng}, \citenamefont {Chen},
  \citenamefont {Wang}, \citenamefont {Mohanty},\ and\ \citenamefont
  {Shao}}]{Feng2023a}%
  \BibitemOpen
  \bibfield  {author} {\bibinfo {author} {\bibfnamefont {W.-F.}\ \bibnamefont
  {Feng}}, \bibinfo {author} {\bibfnamefont {J.-W.}\ \bibnamefont {Chen}},
  \bibinfo {author} {\bibfnamefont {Y.}~\bibnamefont {Wang}}, \bibinfo {author}
  {\bibfnamefont {S.~D.}\ \bibnamefont {Mohanty}}, \ and\ \bibinfo {author}
  {\bibfnamefont {Y.}~\bibnamefont {Shao}},\ }\href {\doibase
  10.1103/PhysRevD.107.103035} {\bibfield  {journal} {\bibinfo  {journal}
  {Phys. Rev. D}\ }\textbf {\bibinfo {volume} {107}},\ \bibinfo {pages}
  {103035} (\bibinfo {year} {2023}{\natexlab{a}})}\BibitemShut {NoStop}%
\bibitem [{\citenamefont {{Punturo}}\ \emph {et~al.}(2010)\citenamefont
  {{Punturo}}, \citenamefont {{Abernathy}},\ and\ \citenamefont
  {et~al.}}]{ET2010}%
  \BibitemOpen
  \bibfield  {author} {\bibinfo {author} {\bibfnamefont {M.}~\bibnamefont
  {{Punturo}}}, \bibinfo {author} {\bibfnamefont {M.}~\bibnamefont
  {{Abernathy}}}, \ and\ \bibinfo {author} {\bibnamefont {et~al.}},\ }\href
  {\doibase 10.1088/0264-9381/27/19/194002} {\bibfield  {journal} {\bibinfo
  {journal} {Class. Quantum Grav.}\ }\textbf {\bibinfo {volume} {27}},\
  \bibinfo {eid} {194002} (\bibinfo {year} {2010})}\BibitemShut {NoStop}%
\bibitem [{\citenamefont {Kalogera}\ \emph {et~al.}(2021)\citenamefont
  {Kalogera} \emph {et~al.}}]{Kalogera:2021bya}%
  \BibitemOpen
  \bibfield  {author} {\bibinfo {author} {\bibfnamefont {V.}~\bibnamefont
  {Kalogera}} \emph {et~al.},\ }\href@noop {} {\  (\bibinfo {year} {2021})},\
  \Eprint {http://arxiv.org/abs/2111.06990} {arXiv:2111.06990 [gr-qc]}
  \BibitemShut {NoStop}%
\bibitem [{\citenamefont {{Srivastava}}\ \emph {et~al.}(2022)\citenamefont
  {{Srivastava}}, \citenamefont {{Davis}}, \citenamefont {{Kuns}},
  \citenamefont {{Landry}}, \citenamefont {{Ballmer}}, \citenamefont {{Evans}},
  \citenamefont {{Hall}}, \citenamefont {{Read}},\ and\ \citenamefont
  {{Sathyaprakash}}}]{CosmicExplorer2022}%
  \BibitemOpen
  \bibfield  {author} {\bibinfo {author} {\bibfnamefont {V.}~\bibnamefont
  {{Srivastava}}}, \bibinfo {author} {\bibfnamefont {D.}~\bibnamefont
  {{Davis}}}, \bibinfo {author} {\bibfnamefont {K.}~\bibnamefont {{Kuns}}},
  \bibinfo {author} {\bibfnamefont {P.}~\bibnamefont {{Landry}}}, \bibinfo
  {author} {\bibfnamefont {S.}~\bibnamefont {{Ballmer}}}, \bibinfo {author}
  {\bibfnamefont {M.}~\bibnamefont {{Evans}}}, \bibinfo {author} {\bibfnamefont
  {E.~D.}\ \bibnamefont {{Hall}}}, \bibinfo {author} {\bibfnamefont
  {J.}~\bibnamefont {{Read}}}, \ and\ \bibinfo {author} {\bibfnamefont {B.~S.}\
  \bibnamefont {{Sathyaprakash}}},\ }\href {\doibase 10.3847/1538-4357/ac5f04}
  {\bibfield  {journal} {\bibinfo  {journal} {\apj}\ }\textbf {\bibinfo
  {volume} {931}},\ \bibinfo {eid} {22} (\bibinfo {year} {2022})},\ \Eprint
  {http://arxiv.org/abs/2201.10668} {arXiv:2201.10668 [gr-qc]} \BibitemShut
  {NoStop}%
\bibitem [{\citenamefont {Poisson}\ and\ \citenamefont
  {Will}(2014)}]{poisson2014gravity}%
  \BibitemOpen
  \bibfield  {author} {\bibinfo {author} {\bibfnamefont {E.}~\bibnamefont
  {Poisson}}\ and\ \bibinfo {author} {\bibfnamefont {C.~M.}\ \bibnamefont
  {Will}},\ }\href@noop {} {\emph {\bibinfo {title} {Gravity: Newtonian,
  post-Newtonian, Relativistic}}}\ (\bibinfo  {publisher} {Cambridge University
  Press},\ \bibinfo {year} {2014})\BibitemShut {NoStop}%
\bibitem [{\citenamefont {Feng}\ \emph
  {et~al.}(2023{\natexlab{b}})\citenamefont {Feng}, \citenamefont {Liu},
  \citenamefont {Chen}, \citenamefont {Wang},\ and\ \citenamefont
  {Mohanty}}]{Feng2023b}%
  \BibitemOpen
  \bibfield  {author} {\bibinfo {author} {\bibfnamefont {W.-F.}\ \bibnamefont
  {Feng}}, \bibinfo {author} {\bibfnamefont {T.}~\bibnamefont {Liu}}, \bibinfo
  {author} {\bibfnamefont {J.-W.}\ \bibnamefont {Chen}}, \bibinfo {author}
  {\bibfnamefont {Y.}~\bibnamefont {Wang}}, \ and\ \bibinfo {author}
  {\bibfnamefont {S.~D.}\ \bibnamefont {Mohanty}},\ }\href {\doibase
  10.1103/PhysRevD.108.063035} {\bibfield  {journal} {\bibinfo  {journal}
  {Phys. Rev. D}\ }\textbf {\bibinfo {volume} {108}},\ \bibinfo {pages}
  {063035} (\bibinfo {year} {2023}{\natexlab{b}})}\BibitemShut {NoStop}%
\bibitem [{\citenamefont {Kremer}\ \emph {et~al.}(2018)\citenamefont {Kremer},
  \citenamefont {Chatterjee}, \citenamefont {Breivik}, \citenamefont
  {Rodriguez}, \citenamefont {Larson},\ and\ \citenamefont
  {Rasio}}]{Kremer2018}%
  \BibitemOpen
  \bibfield  {author} {\bibinfo {author} {\bibfnamefont {K.}~\bibnamefont
  {Kremer}}, \bibinfo {author} {\bibfnamefont {S.}~\bibnamefont {Chatterjee}},
  \bibinfo {author} {\bibfnamefont {K.}~\bibnamefont {Breivik}}, \bibinfo
  {author} {\bibfnamefont {C.~L.}\ \bibnamefont {Rodriguez}}, \bibinfo {author}
  {\bibfnamefont {S.~L.}\ \bibnamefont {Larson}}, \ and\ \bibinfo {author}
  {\bibfnamefont {F.~A.}\ \bibnamefont {Rasio}},\ }\href {\doibase
  10.1103/PhysRevLett.120.191103} {\bibfield  {journal} {\bibinfo  {journal}
  {Phys. Rev. Lett.}\ }\textbf {\bibinfo {volume} {120}},\ \bibinfo {pages}
  {191103} (\bibinfo {year} {2018})}\BibitemShut {NoStop}%
\bibitem [{\citenamefont {Einstein}(1918)}]{Einstein:1918btx}%
  \BibitemOpen
  \bibfield  {author} {\bibinfo {author} {\bibfnamefont {A.}~\bibnamefont
  {Einstein}},\ }\href@noop {} {\bibfield  {journal} {\bibinfo  {journal}
  {Sitzungsber. Preuss. Akad. Wiss. Berlin (Math. Phys.)}\ }\textbf {\bibinfo
  {volume} {1918}},\ \bibinfo {pages} {154} (\bibinfo {year}
  {1918})}\BibitemShut {NoStop}%
\bibitem [{\citenamefont {{Apostolatos}}\ \emph {et~al.}(1994)\citenamefont
  {{Apostolatos}}, \citenamefont {{Cutler}}, \citenamefont {{Sussman}},\ and\
  \citenamefont {{Thorne}}}]{Apostolatos1994}%
  \BibitemOpen
  \bibfield  {author} {\bibinfo {author} {\bibfnamefont {T.~A.}\ \bibnamefont
  {{Apostolatos}}}, \bibinfo {author} {\bibfnamefont {C.}~\bibnamefont
  {{Cutler}}}, \bibinfo {author} {\bibfnamefont {G.~J.}\ \bibnamefont
  {{Sussman}}}, \ and\ \bibinfo {author} {\bibfnamefont {K.~S.}\ \bibnamefont
  {{Thorne}}},\ }\href {\doibase 10.1103/PhysRevD.49.6274} {\bibfield
  {journal} {\bibinfo  {journal} {\prd}\ }\textbf {\bibinfo {volume} {49}},\
  \bibinfo {pages} {6274} (\bibinfo {year} {1994})}\BibitemShut {NoStop}%
\bibitem [{\citenamefont {Barker}\ and\ \citenamefont
  {O'Connell}(1975)}]{Barker1975}%
  \BibitemOpen
  \bibfield  {author} {\bibinfo {author} {\bibfnamefont {B.~M.}\ \bibnamefont
  {Barker}}\ and\ \bibinfo {author} {\bibfnamefont {R.~F.}\ \bibnamefont
  {O'Connell}},\ }\href {\doibase 10.1103/PhysRevD.12.329} {\bibfield
  {journal} {\bibinfo  {journal} {Phys. Rev. D}\ }\textbf {\bibinfo {volume}
  {12}},\ \bibinfo {pages} {329} (\bibinfo {year} {1975})}\BibitemShut
  {NoStop}%
\bibitem [{\citenamefont {Landau}\ and\ \citenamefont
  {Lifshitz}(1976)}]{LANDAU1976}%
  \BibitemOpen
  \bibfield  {author} {\bibinfo {author} {\bibfnamefont {L.}~\bibnamefont
  {Landau}}\ and\ \bibinfo {author} {\bibfnamefont {E.}~\bibnamefont
  {Lifshitz}},\ }in\ \href {\doibase
  https://doi.org/10.1016/B978-0-08-050347-9.50011-3} {\emph {\bibinfo
  {booktitle} {Mechanics (Third Edition)}}}\ (\bibinfo  {publisher}
  {Butterworth-Heinemann},\ \bibinfo {address} {Oxford},\ \bibinfo {year}
  {1976})\ pp.\ \bibinfo {pages} {96--130}\BibitemShut {NoStop}%
\bibitem [{\citenamefont {{Zimmermann}}(1980)}]{Zimmermann1980}%
  \BibitemOpen
  \bibfield  {author} {\bibinfo {author} {\bibfnamefont {M.}~\bibnamefont
  {{Zimmermann}}},\ }\href {\doibase 10.1103/PhysRevD.21.891} {\bibfield
  {journal} {\bibinfo  {journal} {\prd}\ }\textbf {\bibinfo {volume} {21}},\
  \bibinfo {pages} {891} (\bibinfo {year} {1980})}\BibitemShut {NoStop}%
\bibitem [{\citenamefont {{Gao}}\ \emph {et~al.}(2020)\citenamefont {{Gao}},
  \citenamefont {{Shao}}, \citenamefont {{Xu}}, \citenamefont {{Sun}},
  \citenamefont {{Liu}},\ and\ \citenamefont {{Xu}}}]{Gao2020}%
  \BibitemOpen
  \bibfield  {author} {\bibinfo {author} {\bibfnamefont {Y.}~\bibnamefont
  {{Gao}}}, \bibinfo {author} {\bibfnamefont {L.}~\bibnamefont {{Shao}}},
  \bibinfo {author} {\bibfnamefont {R.}~\bibnamefont {{Xu}}}, \bibinfo {author}
  {\bibfnamefont {L.}~\bibnamefont {{Sun}}}, \bibinfo {author} {\bibfnamefont
  {C.}~\bibnamefont {{Liu}}}, \ and\ \bibinfo {author} {\bibfnamefont {R.-X.}\
  \bibnamefont {{Xu}}},\ }\href {\doibase 10.1093/mnras/staa2476} {\bibfield
  {journal} {\bibinfo  {journal} {Mon. Not. Roy. Astron. Soc.}\ }\textbf
  {\bibinfo {volume} {498}},\ \bibinfo {pages} {1826} (\bibinfo {year}
  {2020})},\ \Eprint {http://arxiv.org/abs/2007.02528} {arXiv:2007.02528
  [astro-ph.HE]} \BibitemShut {NoStop}%
\bibitem [{\citenamefont {{Ostriker}}\ and\ \citenamefont
  {{Gunn}}(1969)}]{Ostriker1969}%
  \BibitemOpen
  \bibfield  {author} {\bibinfo {author} {\bibfnamefont {J.~P.}\ \bibnamefont
  {{Ostriker}}}\ and\ \bibinfo {author} {\bibfnamefont {J.~E.}\ \bibnamefont
  {{Gunn}}},\ }\href {\doibase 10.1086/150160} {\bibfield  {journal} {\bibinfo
  {journal} {\apj}\ }\textbf {\bibinfo {volume} {157}},\ \bibinfo {pages}
  {1395} (\bibinfo {year} {1969})}\BibitemShut {NoStop}%
\bibitem [{\citenamefont {Gao}\ \emph {et~al.}(2023)\citenamefont {Gao},
  \citenamefont {Shao}, \citenamefont {Desvignes}, \citenamefont {Jones},
  \citenamefont {Kramer},\ and\ \citenamefont {Yim}}]{Gao:2022hzd}%
  \BibitemOpen
  \bibfield  {author} {\bibinfo {author} {\bibfnamefont {Y.}~\bibnamefont
  {Gao}}, \bibinfo {author} {\bibfnamefont {L.}~\bibnamefont {Shao}}, \bibinfo
  {author} {\bibfnamefont {G.}~\bibnamefont {Desvignes}}, \bibinfo {author}
  {\bibfnamefont {D.~I.}\ \bibnamefont {Jones}}, \bibinfo {author}
  {\bibfnamefont {M.}~\bibnamefont {Kramer}}, \ and\ \bibinfo {author}
  {\bibfnamefont {G.}~\bibnamefont {Yim}},\ }\href {\doibase
  10.1093/mnras/stac3546} {\bibfield  {journal} {\bibinfo  {journal} {Mon. Not.
  Roy. Astron. Soc.}\ }\textbf {\bibinfo {volume} {519}},\ \bibinfo {pages}
  {1080} (\bibinfo {year} {2023})},\ \Eprint {http://arxiv.org/abs/2211.17087}
  {arXiv:2211.17087 [astro-ph.HE]} \BibitemShut {NoStop}%
\bibitem [{\citenamefont {{Van Den Broeck}}(2005)}]{Broeck2005}%
  \BibitemOpen
  \bibfield  {author} {\bibinfo {author} {\bibfnamefont {C.}~\bibnamefont {{Van
  Den Broeck}}},\ }\href {\doibase 10.1088/0264-9381/22/9/022} {\bibfield
  {journal} {\bibinfo  {journal} {Class. Quantum Grav.}\ }\textbf {\bibinfo
  {volume} {22}},\ \bibinfo {pages} {1825} (\bibinfo {year} {2005})},\ \Eprint
  {http://arxiv.org/abs/gr-qc/0411030} {arXiv:gr-qc/0411030 [gr-qc]}
  \BibitemShut {NoStop}%
\bibitem [{\citenamefont {{Apostolatos}}(1995)}]{Apostolatos1995FF}%
  \BibitemOpen
  \bibfield  {author} {\bibinfo {author} {\bibfnamefont {T.~A.}\ \bibnamefont
  {{Apostolatos}}},\ }\href {\doibase 10.1103/PhysRevD.52.605} {\bibfield
  {journal} {\bibinfo  {journal} {\prd}\ }\textbf {\bibinfo {volume} {52}},\
  \bibinfo {pages} {605} (\bibinfo {year} {1995})}\BibitemShut {NoStop}%
\bibitem [{\citenamefont {Mukherjee}\ \emph {et~al.}(2023)\citenamefont
  {Mukherjee}, \citenamefont {Prix},\ and\ \citenamefont
  {Wette}}]{Mukherjee:2022tuc}%
  \BibitemOpen
  \bibfield  {author} {\bibinfo {author} {\bibfnamefont {A.}~\bibnamefont
  {Mukherjee}}, \bibinfo {author} {\bibfnamefont {R.}~\bibnamefont {Prix}}, \
  and\ \bibinfo {author} {\bibfnamefont {K.}~\bibnamefont {Wette}},\ }\href
  {\doibase 10.1103/PhysRevD.107.062005} {\bibfield  {journal} {\bibinfo
  {journal} {Phys. Rev. D}\ }\textbf {\bibinfo {volume} {107}},\ \bibinfo
  {pages} {062005} (\bibinfo {year} {2023})},\ \Eprint
  {http://arxiv.org/abs/2207.09326} {arXiv:2207.09326 [gr-qc]} \BibitemShut
  {NoStop}%
\bibitem [{\citenamefont {Hulse}\ and\ \citenamefont
  {Taylor}(1975)}]{HulseTaylor1975}%
  \BibitemOpen
  \bibfield  {author} {\bibinfo {author} {\bibfnamefont {R.~A.}\ \bibnamefont
  {Hulse}}\ and\ \bibinfo {author} {\bibfnamefont {J.~H.}\ \bibnamefont
  {Taylor}},\ }\href {\doibase 10.1086/181708} {\bibfield  {journal} {\bibinfo
  {journal} {Astrophys. J. Lett.}\ }\textbf {\bibinfo {volume} {195}},\
  \bibinfo {pages} {L51} (\bibinfo {year} {1975})}\BibitemShut {NoStop}%
\bibitem [{\citenamefont {Zhao}\ \emph {et~al.}(2021)\citenamefont {Zhao},
  \citenamefont {Shao}, \citenamefont {Gao}, \citenamefont {Liu}, \citenamefont
  {Cao},\ and\ \citenamefont {Ma}}]{Zhao:2021bjw}%
  \BibitemOpen
  \bibfield  {author} {\bibinfo {author} {\bibfnamefont {J.}~\bibnamefont
  {Zhao}}, \bibinfo {author} {\bibfnamefont {L.}~\bibnamefont {Shao}}, \bibinfo
  {author} {\bibfnamefont {Y.}~\bibnamefont {Gao}}, \bibinfo {author}
  {\bibfnamefont {C.}~\bibnamefont {Liu}}, \bibinfo {author} {\bibfnamefont
  {Z.}~\bibnamefont {Cao}}, \ and\ \bibinfo {author} {\bibfnamefont {B.-Q.}\
  \bibnamefont {Ma}},\ }\href {\doibase 10.1103/PhysRevD.104.084008} {\bibfield
   {journal} {\bibinfo  {journal} {Phys. Rev. D}\ }\textbf {\bibinfo {volume}
  {104}},\ \bibinfo {pages} {084008} (\bibinfo {year} {2021})},\ \Eprint
  {http://arxiv.org/abs/2106.04883} {arXiv:2106.04883 [gr-qc]} \BibitemShut
  {NoStop}%
\bibitem [{\citenamefont {Riley}\ \emph {et~al.}(2022)\citenamefont {Riley}
  \emph {et~al.}}]{COMPAS}%
  \BibitemOpen
  \bibfield  {author} {\bibinfo {author} {\bibfnamefont {J.}~\bibnamefont
  {Riley}} \emph {et~al.} (\bibinfo {collaboration} {COMPAS Team, Team
  COMPAS}),\ }\href {\doibase 10.3847/1538-4365/ac416c} {\bibfield  {journal}
  {\bibinfo  {journal} {Astrophys. J. Supp.}\ }\textbf {\bibinfo {volume}
  {258}},\ \bibinfo {pages} {34} (\bibinfo {year} {2022})},\ \Eprint
  {http://arxiv.org/abs/2109.10352} {arXiv:2109.10352 [astro-ph.IM]}
  \BibitemShut {NoStop}%
\bibitem [{\citenamefont {{Jaranowski}}\ \emph {et~al.}(1998)\citenamefont
  {{Jaranowski}}, \citenamefont {{Krolak}},\ and\ \citenamefont
  {{Schutz}}}]{Jaranowski1998}%
  \BibitemOpen
  \bibfield  {author} {\bibinfo {author} {\bibfnamefont {P.}~\bibnamefont
  {{Jaranowski}}}, \bibinfo {author} {\bibfnamefont {A.}~\bibnamefont
  {{Krolak}}}, \ and\ \bibinfo {author} {\bibfnamefont {B.~F.}\ \bibnamefont
  {{Schutz}}},\ }\href {\doibase 10.1103/PhysRevD.58.063001} {\bibfield
  {journal} {\bibinfo  {journal} {\prd}\ }\textbf {\bibinfo {volume} {58}},\
  \bibinfo {eid} {063001} (\bibinfo {year} {1998})},\ \Eprint
  {http://arxiv.org/abs/gr-qc/9804014} {arXiv:gr-qc/9804014 [gr-qc]}
  \BibitemShut {NoStop}%
\bibitem [{\citenamefont {{Jaranowski}}\ and\ \citenamefont
  {{Kr{\'o}lak}}(1999)}]{Jaranowski1999PhRvD}%
  \BibitemOpen
  \bibfield  {author} {\bibinfo {author} {\bibfnamefont {P.}~\bibnamefont
  {{Jaranowski}}}\ and\ \bibinfo {author} {\bibfnamefont {A.}~\bibnamefont
  {{Kr{\'o}lak}}},\ }\href {\doibase 10.1103/PhysRevD.59.063003} {\bibfield
  {journal} {\bibinfo  {journal} {\prd}\ }\textbf {\bibinfo {volume} {59}},\
  \bibinfo {eid} {063003} (\bibinfo {year} {1999})},\ \Eprint
  {http://arxiv.org/abs/gr-qc/9809046} {arXiv:gr-qc/9809046 [gr-qc]}
  \BibitemShut {NoStop}%
\bibitem [{\citenamefont {Shah}\ \emph {et~al.}(2012)\citenamefont {Shah},
  \citenamefont {van~der Sluys},\ and\ \citenamefont {Nelemans}}]{Shah2012}%
  \BibitemOpen
  \bibfield  {author} {\bibinfo {author} {\bibfnamefont {S.}~\bibnamefont
  {Shah}}, \bibinfo {author} {\bibfnamefont {M.}~\bibnamefont {van~der Sluys}},
  \ and\ \bibinfo {author} {\bibfnamefont {G.}~\bibnamefont {Nelemans}},\
  }\href {\doibase 10.1051/0004-6361/201219309} {\bibfield  {journal} {\bibinfo
   {journal} {Astron. Astrophys.}\ }\textbf {\bibinfo {volume} {544}},\
  \bibinfo {pages} {A153} (\bibinfo {year} {2012})},\ \Eprint
  {http://arxiv.org/abs/1207.6770} {arXiv:1207.6770 [astro-ph.IM]} \BibitemShut
  {NoStop}%
\bibitem [{\citenamefont {Cutler}\ and\ \citenamefont
  {Vecchio}(1998)}]{cutler1998lisa}%
  \BibitemOpen
  \bibfield  {author} {\bibinfo {author} {\bibfnamefont {C.}~\bibnamefont
  {Cutler}}\ and\ \bibinfo {author} {\bibfnamefont {A.}~\bibnamefont
  {Vecchio}},\ }in\ \href@noop {} {\emph {\bibinfo {booktitle} {AIP Conference
  Proceedings}}},\ Vol.\ \bibinfo {volume} {456}\ (\bibinfo {organization}
  {American Institute of Physics},\ \bibinfo {year} {1998})\ pp.\ \bibinfo
  {pages} {95--100}\BibitemShut {NoStop}%
\end{thebibliography}%

\end{document}